\documentclass[fleqn,usenatbib]{mnras}

\usepackage{newtxtext,newtxmath}

\usepackage[T1]{fontenc}
\usepackage{ae,aecompl}

\usepackage{graphicx}	
\usepackage{amsmath}	
\usepackage{amssymb}	

\def\lya{{\rm\,Ly$\alpha$}}
\def\ha{{\rm\,H$\alpha$}}
\def\hb{{\rm\,H$\beta$}}

\def\oii{{\rm\,[O{\sc ii}]}}
\def\nii{{\rm\,[N{\sc ii}]}}

\def\siii{{\rm\,[S{\sc iii}]}}
\def\oiii{{\rm\,[O{\sc iii}]}}

\def\hi{{\rm\,H{\sc i}}}

\def\msun{{\rm M}$_{\odot}$}

\title[MDCS--I. Galaxy formation in the densest regions]
{MAHALO Deep Cluster Survey I. Accelerated and enhanced galaxy formation in the densest regions of a protocluster at $z=2.5$}

\author[R. Shimakawa et al.]{
Rhythm Shimakawa,$^{1}$\thanks{rhythm@ucolick.org} 
Tadayuki Kodama,$^{2}$ 
Masao Hayashi,$^{3}$
J. Xavier Prochaska,$^{1}$
\newauthor
Ichi Tanaka,$^{4}$
Zheng Cai,$^{1}$
Tomoko L. Suzuki,$^{3}$
Ken-ichi Tadaki$^{3}$
and Yusei Koyama$^{4}$
\\
$^{1}$UCO/Lick Observatory, University of California, 1156 High Street, Santa Cruz, CA 95064, USA\\
$^{2}$Astronomical Institute, Tohoku University, Aoba-ku, Sendai 980-8578, Japan\\
$^{3}$National Astronomical Observatory of Japan, Osawa, Mitaka, Tokyo 181-8588, Japan\\
$^{4}$Subaru Telescope, National Astronomical Observatory of Japan, National Institutes of Natural Sciences, 650 North A'ohoku Place, Hilo, HI 96720, USA\\
}

\date{Accepted XXX. Received YYY; in original form 2017 August 21}

\pubyear{2017}

\begin{document}
\label{firstpage}
\pagerange{\pageref{firstpage}--\pageref{lastpage}}
\maketitle

\begin{abstract}
We carried out deep H$\alpha$ narrowband imaging with 10 hours net integrations 
towards the young protocluster, USS1558$-$003 at $z=2.53$ with the Subaru 
Telescope. This system is composed of four dense groups with massive local 
overdensities, traced by 107 H$\alpha$ emitters (HAEs) with stellar masses and 
dust-corrected star formation rates down to $1\times10^8$ M$_\odot$ and 3 
M$_\odot$yr$^{-1}$, respectively. We have investigated the environmental dependence 
of various physical 
properties within the protocluster by comparing distributions of HAEs in higher and 
lower densities with a standard Kolmogorov--Smirnov test. At 97\% confidence level,
we find enhanced star formation across the star-forming main sequence of HAEs 
living in the most extreme `supergroup', corresponding to the top quartile of 
overdensities. Furthermore, we derive distribution functions of H$\alpha$ 
luminosity and stellar mass in group and intergroup regions, approximately 
corresponding to 30 times and 8 times higher densities than the general field. As a 
consequence, we identify 0.7 and 0.9 dex higher cutoffs in H$\alpha$ luminosity 
and stellar mass functions in the dense groups, respectively. On the other hand, 
HAEs in the intergroup environment of the protocluster show similar distribution 
functions to those of field galaxies despite residing in significant overdensities. 
In the early phase of cluster formation, as inferred from our results, the densest 
parts in the protocluster have had an accelerated formation of massive galaxies. We 
expect that these eventually grow and transform into early-type galaxies at the 
bright end of the red sequence as seen in present-day rich clusters of galaxies.
\end{abstract}




\section{Introduction}\label{s1}

Galaxy protoclusters\footnotemark[1] \citep{Albada:1961,Peebles:1970a,Sunyaev:1972} 
provide us with direct insight into the early phase of hierarchical inside-out 
growth of the galaxy clusters. Studying these unique laboratories also leads deep 
understanding such as formation histories of strong colour--magnitude relationships 
as known as the red sequence 
\citep{Vaucouleurs:1961,Visvanathan:1977,Butcher:1984,Bower:1992,Bower:1998,Terlevich:2001,Tanaka:2005,Kodama:2007,Mei:2009} 
and physical origins of galaxy segregation i.e. morphology--density relation 
as seen in the local galaxy clusters 
(e.g. \citealt{Dressler:1980,Dressler:1997,Couch:1998,Goto:2003}, and more modern 
examples by \citealt{Cappellari:2011,Houghton:2013,Fogarty:2014,Brough:2017}). 

\footnotetext[1]{Various survey bias and restrictions result in vague and 
inconsistent definitions of the {\it protocluster} in any work. This series of 
papers refers to overdense regions on the spatial scale of $\gtrsim10$, $\sim$1--10, 
and $\lesssim1$ comoving Mpc as large-scale structures, protoclusters, and dense 
groups (or cores) for the target, respectively.}

Protoclusters at redshifts greater than two have been extensively searched for and 
studied over the past two decades (e.g. 
\citealt{Steidel:1998,Kurk:2000,Kurk:2004a,Venemans:2002,Mayo:2012,Shimakawa:2014,Planck:2015,Toshikawa:2016,Cai:2016,Toshikawa:2017}). 
The results and conclusions from the past studies are not always consistent, but 
still, these do not have to be coherent with each other since they are based on 
different protoclusters with various overdensities, scale, sample selection, and 
redshifts (e.g. conflicts in gaseous metallicities of protocluster galaxies: 
\citealt{Kulas:2013,Shimakawa:2015,Valentino:2015,Kacprzak:2015,Tran:2015}). 
As a general tendency, however, one tends to confirm a larger number of massive 
and/or red star-forming galaxies in massive protoclusters as compared to the 
general field 
\citep{Steidel:2005,Hatch:2011,Matsuda:2011b,Hayashi:2012,Koyama:2013a,Koyama:2013b,Wang:2016,Husband:2016,Hatch:2017,Ao:2017,Oteo:2017}. 
This implies that rapid and early build-up of bright red sequence objects has 
occurred in a young phase of protocluster formation. Indeed, massive clusters of 
galaxies at lower-$z$ host a higher fraction of very massive quiescent galaxies 
than lower-density environments, just like {\it skyscrapers in the big city} 
\citep{Balogh:2001,Bamford:2009,Blanton:2009,Muzzin:2012,Darvish:2016,Tomczak:2017}. 

In the local Universe, \citet{Bamford:2009} and \citet{Skibba:2009} reported that, 
based on data from the Galaxy Zoo project \citep{Lintott:2008,Lintott:2011}, 
the strong colour dependence remains prominent at fixed stellar mass in the sense 
that the high-density environments tend to have a higher fraction of red galaxies 
(see also \citealt{Zehavi:2011,Kelkar:2017}). 
This trend is particularly decisive at the lower mass regime, suggesting that 
local overdensities would control those quenching histories more efficiently 
\citep{Peng:2010,Thomas:2010,Bolzonella:2010,Mortlock:2015}. 
More interestingly, however, the (kinematic) morphology--density relation is much 
less prominent when one compares galaxies across density environments for a given 
stellar mass range \citep{Bamford:2009,Brough:2017}. 
\citet{Bamford:2009,Brough:2017} argue that such morphological trends are primarily 
driven by stellar mass rather than environments, while the larger number of very 
massive galaxies in clusters of galaxies can give an apparent morphology--density 
relation in these systems. 

Due to this background, high-$z$ protoclusters are ideal testbeds for understanding 
the physical processes that drive substantial excess of bright red sequence 
galaxies in today's cluster centres since these massive galaxies are expected to be 
built in the early stage of cluster formation at $z>2$ (e.g. 
\citealt{Kodama:2007,Zirm:2008,Snyder:2012,Strazzullo:2016}, and see also 
theoretical work by e.g. 
\citealt{Romeo:2005,Jimenez:2011,Gabor:2012,Behroozi:2013b}). 
However, early environmental dependence of massive galaxy formation remains 
controversial based solely on the perspective of analyses of passive galaxies in 
clusters or protoclusters 
\citep{Thomas:2005,Thomas:2010,Strazzullo:2013,Gobat:2013,Tanaka:2013b,Newman:2014b,Mateu:2014}. 
To seek a clearer picture, we require a more direct investigations towards 
the massive {\it forming} galaxies in the young protoclusters at high redshifts 
where star formation is still actively ongoing. 

With this motivation, addressing the stellar mass function of forming galaxies in 
the early phase of protoclusters is essential to identify the physical origins of 
the environmental dependence of the stellar mass function for early type galaxies 
in today's galaxy clusters. Also, those star formation and its distribution 
provide us with original insight into the formation histories of massive red 
sequence galaxies. Toward these particular goals, we have carried out the MAHALO 
Deep Custer Survey (MDCS) as an extension of our previous systematic campaign 
the Mapping H-Alpha and Lines of Oxygen with Subaru (MAHALO-Subaru, 
\citealt{Kodama:2013}). Both projects are designed to search and map \ha\ or \oii\ 
line-emitting galaxies associated with distant clusters, protoclusters, and random
fields at certain redshift slices traced by our narrowband filters. The narrowband 
filters are used with the Subaru Prime Focus Camera (Suprime-Cam; 
\citealt{Miyazaki:2002}) or the Multi-Object InfraRed Camera and Spectrograph 
(MOIRCS; \citealt{Ichikawa:2006,Suzuki:2008}) on the Subaru Telescope. The MCDS aims 
to get deeper \ha\ imaging data for two dense protocluster regions among our parent 
sample, PKS~1138$-$262 at $z=2.2$ \citep{Koyama:2013a} and USS~1558$-$003 at $z=2.5$ 
\citep{Hayashi:2012} with net integrations amounting to a total of 5 hrs and 
10 hrs, respectively, which are extended from exposure time of 3 hrs for each by 
MAHALO-Subaru. 
The extended sample resolves spatial distributions of galaxies on the scale of 
$<1$ physical Mpc that resolves substructures within the protoclusters. High sampling 
density is quite essential for the study of such high-$z$ protoclusters since those 
structure formation would not be steady down yet, and spatial distributions of 
protocluster members are expected to be more scattered and clumpy 
\citep{Muldrew:2015,Chiang:2017}. Furthermore, 
a larger sample allows us to more easily determine the characteristic 
\ha\ luminosity ($L_\mathrm{H\alpha}^\ast$) and stellar mass ($M_\star^\ast$) on 
these distribution functions and may also provides insight into those faint-end 
profiles (i.e. the slope of power law, $\alpha$). These will help us to know if local 
environment can drive stellar mass assembly histories in an early phase of cluster 
formation at high redshift.

This work represents the full results obtained from the 10 hrs deep narrowband 
imaging for USS~1558$-$003 at $z=2.5$ followed by our initial report by 
\citet{Hayashi:2016}. Our analyses improve sample selection (\S\ref{s2}). We then 
compare various physical properties and \ha\ luminosity and stellar mass functions 
of obtained sample across local overdensities (\S\ref{s3}). Section \S\ref{s4} 
describes discussions on similarities and differences between HAEs in higher and 
lower densities, and we summarise them in the final part (\S\ref{s5}). This paper 
assumes the cosmological parameters of $\Omega_M=0.3$, $\Omega_\Lambda=0.7$ and 
$h=0.7$ and employs a \citet{Chabrier:2003} stellar initial mass function. The AB 
magnitude system \citep{Oke:1983} are used throughout the paper.


\section{Target and dataset}\label{s2}


\subsection{USS~1558$-$003}\label{s2.1}

USS~1558$-$003 
($\alpha_\mathrm{J2000}=$ 16$^\mathrm{h}$01$^\mathrm{m}$17$^\mathrm{s}$, 
$\delta_\mathrm{J2000}=$ $-$00$^\mathrm{d}$28$^\mathrm{m}$47$^\mathrm{s}$, 
hereafter USS~1558) is located at $z=2.53$, the highest redshift where the \ha\ 
emission line is observable from the ground. USS~1558 was first discovered as an 
overdensity by a factor of four in red galaxies associated with the radio galaxy 
(RG) USS~1558$-$003 \citep{Kajisawa:2006}, but without a prominent red sequence 
\citep{Kodama:2007,Galametz:2012}. Instead, we see four overdense groups associated 
with \ha-emitting galaxies \citep{Hayashi:2012,Hayashi:2016}. 
These suggest that this protocluster system is a young, unrelaxed phase in contrast 
to more advanced cluster systems at similar redshifts. Observationally, USS~1558 
is a valuable target for studying young protocluster systems since this is at a 
redshift supported by \ha\ line and brighter than more distant protoclusters. 

The density peak of USS~1558 corresponds to an approximately 30 times higher number 
density of \ha-selected star-forming galaxies relative to the general field 
\citep{Hayashi:2012}. This object is the densest {\it star-forming} protocluster 
discovered to date at $z>2$ compared with other familiar massive protoclusters such 
as PKS~1138$-$262 \citep{Kurk:2000,Koyama:2013a}, 4C23.56 at $z=2.5$ 
\citep{Knopp:1997,Tanaka:2011}, HS~1700+643 at $z=2.3$ 
\citep{Steidel:2005,Erb:2011}, 2QZ at $z=2.3$ \citep{Matsuda:2011b,Kato:2016}, 
and Cl~J1449+0856 at $z=2.0$ \citep{Gobat:2011,Gobat:2013}. The overdensity is 
such that the mean projected separation of HAEs in the density peak is smaller 
than 100 kpc (\S\ref{s3.1}). Thus, it is highly likely that their sub-haloes begin 
to overlap one another common group halo. 

Four dense groups of HAEs \citep{Hayashi:2016} align from the northwest to the 
southeast (NW--SE) direction: a small group located in the north of the RG, a 
loose group around the RG and six distant red galaxies without spectroscopic 
confirmation \citep{Hayashi:2012}, the richest clump at 3.5 arcmin away from the RG 
containing 25 HAEs within just 1 arcmin$^2$, and a small clump in between the two. 
The spatial distribution is also available in Fig.~\ref{fig3} of this paper. These 
structures are well aligned along the line of sight according to existing 
spectroscopic data \citep{Shimakawa:2014}. The dense groups are likely to 
eventually merge and become a single rich galaxy cluster system. Although we do not 
know whether these groups are virialised, the dynamical masses are estimated to be 
$\sim1\times10^{14}$ \msun\ for the densest group region when we assume local 
virialisation \citep{Shimakawa:2014} according to the \citet{Finn:2005} 
prescription based on the measured velocity dispersion. Such a massive overdensity 
may grow into today's most massive clusters like Coma \citep{Chiang:2013}.


\subsection{Data}\label{s2.2}

\begin{figure}
\centering
\includegraphics[width=0.9\columnwidth]{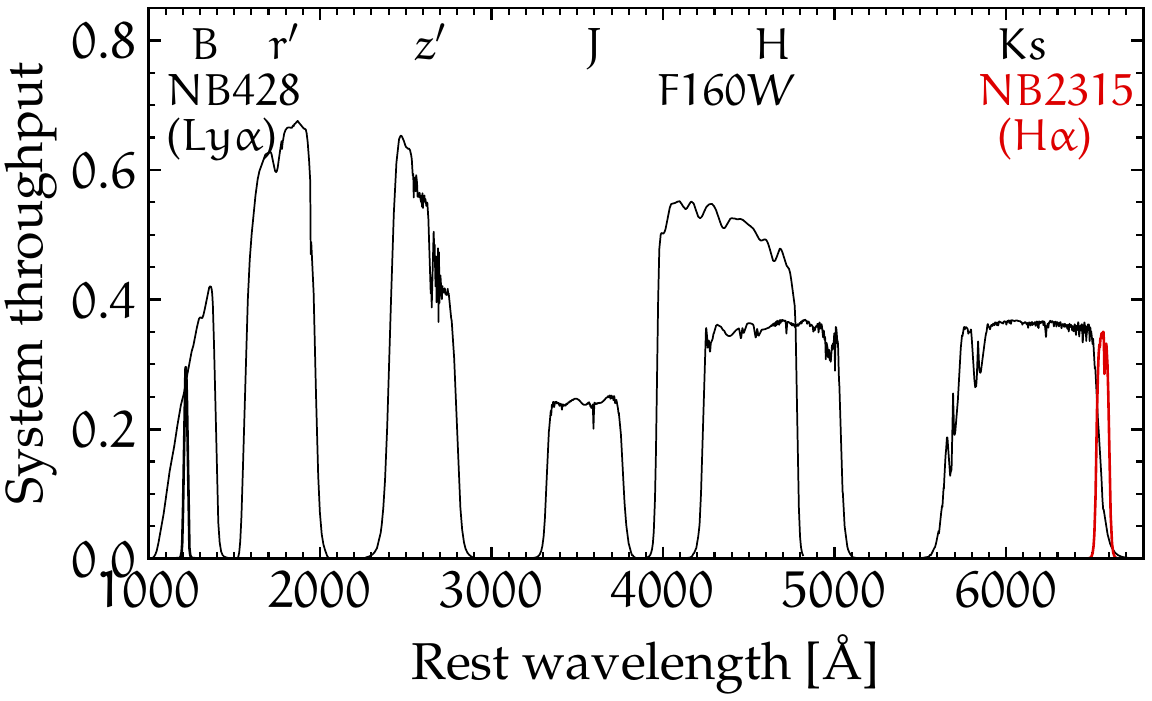}
\caption{System throughputs of available photometric band data (Table~\ref{tab1}) 
for the USS~1558 protocluster field. Rest wavelength is based on the protocluster 
redshift $z=2.53$. The $NB2315$ filter has a companion filter, $NB428$ for the 
\lya\ line at the same redshift slice (see 
\citealt{Shimakawa:2017a,Shimakawa:2017b}). This paper uses $NB428$ photometry only 
to select secure HAE members showing both \ha\ and \lya\ flux excesses in the pair 
of narrowband filters (see text).}
\label{fig1}
\end{figure}

This work combines the multi-broadband and deep narrowband images (Fig.~\ref{fig1}) 
taken from the early data of the MAHALO-Subaru campaign (S10B-028I, Kodama et al.; 
\citealt{Hayashi:2012,Kodama:2013}), the new data by MAHALO Deep Cluster Survey 
(MDCS: S15A-047, Kodama et al.; \citealt{Hayashi:2016}), and deep $F160W$/WFC3 
photometry taken by the Hubble Space Telescope (HST). Although the data are already 
described by \citet{Hayashi:2016}, we summarise the observational details for the 
recent deep follow-up narrowband and broadband imaging with MOIRCS on the Subaru 
Telescope through the MDCS program.

In this observing run, we mainly focused on narrowband imaging with $NB2315$ 
($\lambda_\mathrm{center}=$ 2.317 $\mu$m, FWHM= 260 \AA) that can capture the \ha\ 
emission line from galaxies associated with USS~1558 at $z=2.53\pm0.02$. 
Observations were carried out during April 30 and May 6, 2015, under photometric 
conditions and with seeing of FWHM $\lesssim0.6$ arcsec. The integration time 
of 380 min split to 180 sec individual exposures. Furthermore, we took $J$ 
(116 min) and $K_s$ (150 min) imaging. Combined with our previous data, this 
observing run has reconstructed those photometric data by total integration times 
of 191 min for $J$, 207 min for $K_s$, and 583 min for $NB2315$, respectively. 

Reconstructed $J, K_s, NB2315$ data were reduced very carefully by using the 
reduction pipeline {\sc mcsred}\footnotemark[2] \citep{Tanaka:2011}. {\sc mcsred} 
is composed of {\sc iraf}\footnotemark[3] script. The reduction procedures include 
flat fielding, masking objects from the combined data in the 1st run (thus the 
whole reduction process was conducted twice to remake secure object masks), sky 
subtraction (by median sky and then a polynomially-fitted surface for residual sky 
subtraction), distortion correction, and image mosaicing. Those processes can be 
executed semi-automatically. Throughout these processes, we eventually 
reconstructed deeper $J$, $K_s$, and $NB2315$ data. 

\footnotetext[2]{\url{http://www.naoj.org/staff/ichi/MCSRED/mcsred.html}}
\footnotetext[3]{\url{http://iraf.noao.edu}}

Taken together, $B$, $r'$, $z'$, $J$, $H$, $F160W$, $K_s$, and $NB2315$ data are 
available in USS~1558. Those filter system throughputs are shown in 
Figure~\ref{fig1}. Table~\ref{tab1} describes a brief overview of PSF-matched data 
where Galactic extinctions are derived from the NASA Extragalactic Database 
extinction law calculator\footnotemark[4] which is based on the 
\citet{Fitzpatrick:1999} extinction law under an assumption of $R_V$ = 3.1 and 
\citet{Schlegel:1998,Schlafly:2011} for the dust reddening. We adopt 
E(B$-$V) = 0.1339 mag by \citet{Schlafly:2011}. 

\footnotetext[4]{\url{http://irsa.ipac.caltech.edu/applications/DUST/}}

\begin{table*}
\centering
\caption{Data summary (the same as Table~1 in \citealt{Hayashi:2016}). The first to 
seventh columns indicate filter name, instrument/telescope, filter centre wavelength, 
and filter FWHM, $5\sigma$ limiting magnitude in 1"2 aperture diameter, and galactic 
extinction, respectively. The eighth column shows the references for each data.}
\begin{tabular}{lccccccc}
\hline
Filter & Instrument/Telescope & $\mathrm{\lambda_{center}}$ & Integration & FWHM     & 5$\sigma$ & A$\mathrm{_\lambda}$ & Reference \\
 &   & ($\mathrm{\mu}$m)           & (min)       & (arcsec) & (AB)      & (mag)    &            \\
\hline
$NB2315$& MOIRCS/Subaru & 2.32 & 583 & 0.67 & 23.90 & 0.05 & \citet{Hayashi:2012,Hayashi:2016} \\
$K_s$ & MOIRCS/Subaru & 2.15 & 207 & 0.67 & 24.49 & 0.05 & \citet{Hayashi:2012,Hayashi:2016} \\
$H$   & MOIRCS/Subaru & 1.64 &  45 & 0.67 & 23.78 & 0.07 & \citet{Hayashi:2012,Hayashi:2016} \\
$F160W$& WFC3/HST     & 1.54 &  87 & 0.21 & 27.46 & 0.08 & \citet{Hayashi:2012,Hayashi:2016} \\
$J$   & MOIRCS/Subaru& 1.25 & 191 & 0.67 & 24.85 & 0.11 & \citet{Hayashi:2012,Hayashi:2016} \\
$z'$  & Suprime-Cam/Subaru& 0.91 &  55 & 0.67 & 26.03 & 0.19 & \citet{Hayashi:2012} \\
$r'$  & Suprime-Cam/Subaru& 0.63 &  90 & 0.67 & 27.24 & 0.35 & \citet{Hayashi:2012} \\
$B$   & Suprime-Cam/Subaru& 0.45 &  80 & 0.70 & 27.51 & 0.55 & \citet{Hayashi:2012} \\
\hline
\end{tabular}
\label{tab1}
\end{table*}


\subsection{Sample selection}\label{s2.3}

This work is based on the narrowband emitter sample of \citet{Hayashi:2016}. 
\citet{Hayashi:2016} have selected the sample as these sources showing more than 
$3\sigma$ flux excess at $NB2315$ relative to $K_s$ band photometry in the 
following criteria, 
\begin{eqnarray}
Ks-NB &>& -2.5\log\bigg(1-\frac{\sqrt{f_{3\sigma,Ks} +f_{3\sigma,NB}}}{f_{NB}}\bigg) +0.1 \label{eq1} \\
Ks-NB &>& 0.35 \label{eq2} 
\end{eqnarray}
where $f_{NB}$, $f_{3\sigma,Ks}$, and $f_{3\sigma,NB}$ are 
narrowband flux density, $3\sigma$ limit of $K_s$-band flux density, and $3\sigma$ 
limit of narrowband flux density, respectively. 
We assumed two sigma limiting magnitude in $K_s$-band for faint 
$K_s$ sources, which amount to 26 objects in 177 selected narrowband emitters 
\citep{Hayashi:2016}.
The $3\sigma$ flux limit is 
$1\times10^{-17}$ erg~s$^{-1}$cm$^{-2}$, corresponding to a SFR limit of 2.4 
M$_\odot$yr$^{-1}$ for $z=2.53$ assuming the the \citet{Kennicutt:1998} 
prescription. The latter colour cut corresponds to the \ha\ line equivalent width 
(EW) limit of 45 \AA\ in the rest frame at $z=2.53$. 
One should note that the NB2315 filter is located at the red end of 
$K_s$-band, which requires a colour-term correction for \ha\ line flux and EW 
measurements. This may also lead us to overestimate \ha\ fluxes in redder objects. 
We carefully checked and corrected this effect (\S\ref{s3.2} and Appendix~\ref{a2}) 
and find that it should have a negligible impact on our results and conclusions.

\begin{figure}
\centering
\includegraphics[width=0.9\columnwidth]{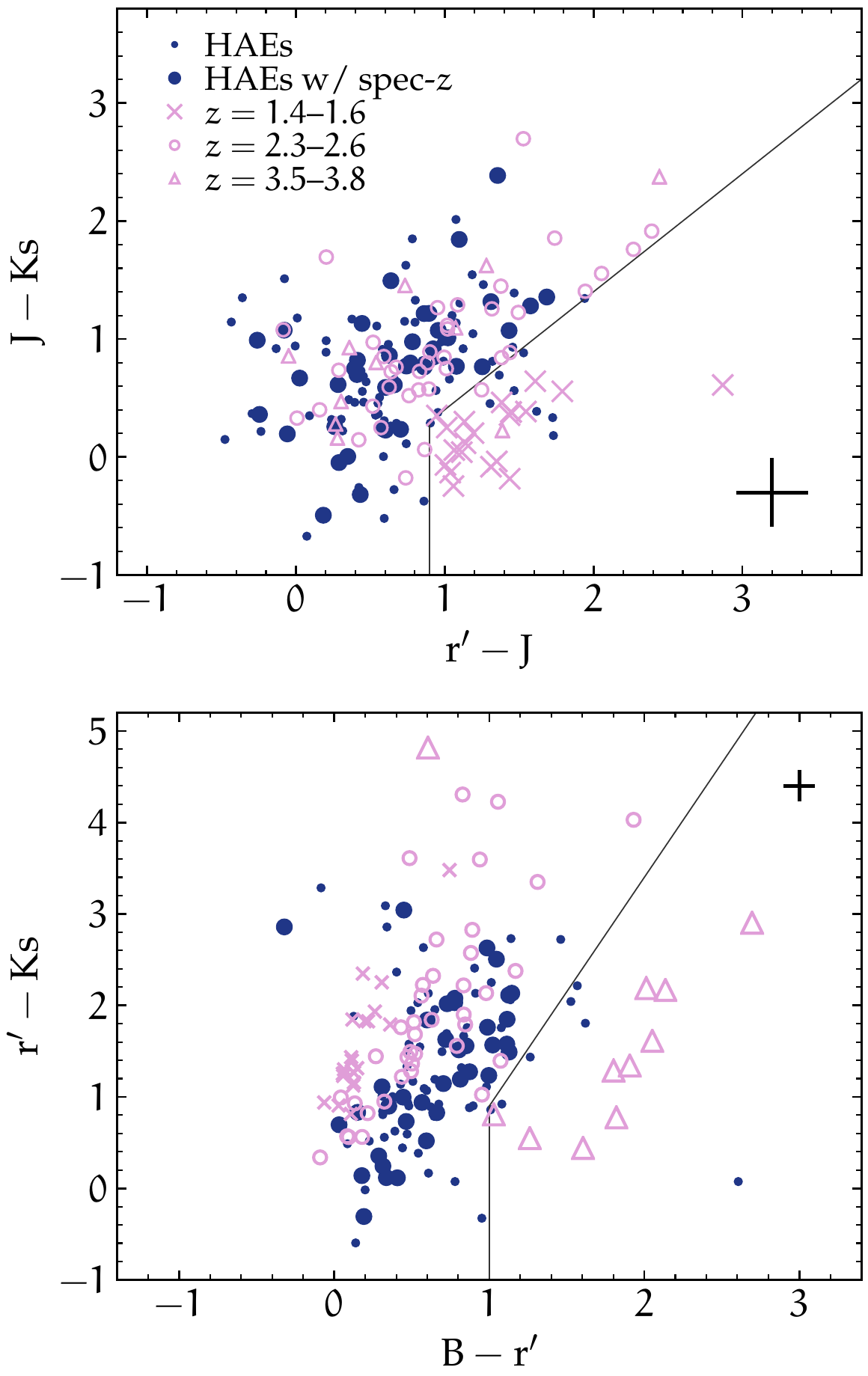}
\caption{$r'JK_s$ ({\it upper panel}) and $Br'K_s$ ({\it lower panel}) 
colour--colour selections for USS~1558. Large and small blue filled circles 
indicate the narrowband emitters with and without confirmation by spectroscopy 
or a high \lya\ EW, respectively. Open pink triangles, circles, and crosses are 
the spec-$z$ samples at $z=3.0-3.3$. $z=2.3-2.6$, and $z=1.2-1.4$ from the MOSDEF 
survey \citep{Kriek:2015}, respectively. Black lines are colour thresholds 
defined to remove those foreground or background contaminants. The black 
crosses on each panel show typical $1\sigma$ errors of our narrowband 
emitters. One should note that only galaxies with better than $2\sigma$ detections 
in all bands appear in these panels.}
\label{fig2}
\end{figure}

A subset of our HAE sample will be emitters of other emission lines in the 
foreground and background. 
According to \citet{Sobral:2013} and \citet{An:2014}, 
\ha\ and \oiii\ emitters would consist mostly of K-band selected emitters, which 
account for approximately 40--50 and 20--30 per cents of the entire emitter sample 
in the general fields, respectively.
To restrict the sample to a more likely set of protocluster members, we need to 
conduct further analyses. We first selected secure protocluster members which were 
confirmed by previous follow-up spectroscopy \citep{Shimakawa:2014,Shimakawa:2015b} 
or \lya\ narrowband imaging \citep{Shimakawa:2017b}. These confirmed HAEs amount to 
a total of 49 objects (Table~\ref{tab2}). \citet{Hayashi:2016} carried out further 
colour--colour selection i.e. $r'JKs$ and $r'H_{F160W}K_s$ to select HAEs at 
$z=2.53\pm0.02$ for narrowband emitters without such identifications. We revisit 
these colour selection criteria to select protocluster members more effectively, and 
finally revise them by new rules on $r'JK_s$ and $Br'K_s$ colour--colour selections 
with the help of spectroscopically confirmed members and spec-$z$ sources by MOSDEF 
survey (the MOSFIRE Deep Evolution Field survey, second redshift release by 
\citealt{Kriek:2015}). In the latter sample, we employ only galaxies in the 
COSMOS-CANDELS field, which allow us to use the same broadband datasets as for 
USS~1558 based on the 3D-HST database \citep{Brammer:2012,Skelton:2014}. The 
colour--colour selections are shown in Figure~\ref{fig2}, and the selection criteria 
are defined as follows,
\begin{eqnarray}
r'-J < 0.9 &or& (J-Ks) > (r'-J) -0.6 \label{eq3} \\
B-r' < 1.0 &or& (r'-Ks) > 2.5\times(B-r') -1.6. \label{eq4}
\end{eqnarray}
$r'JK_s$ can be used to remove low-$z$ contaminants at $z<1.6$ as demonstrated in 
$Bz'K_s$ selections \citep{Daddi:2004,Sobral:2013}. We here use this selection to 
mainly remove \siii$\lambda\lambda$9071,9533 emitters at $z=$ 1.4--1.6 which may be 
one of the major contaminations according to \citet{An:2014}. Then, $Br'K_s$ 
colour cut is useful to reject background \oiii$\lambda\lambda$4960,5008 emitters 
at $z\sim3.6$. Since those Lyman break features stride across between $B$ and 
$r'$-band wavelengths, their $B-r'$ colours are distinctly redder than those of 
galaxies at $z\sim2$. In this colour selection, our sample at $z=2.5$ tends to have 
redder $B-r'$ colours than the MOSDEF sources at the typical spectroscopic redshift 
of $z=2.3$ because those of our HAEs are closer to the Lyman break. 

These colour criteria allow us to select an additional 58 HAEs and remove a total 
of 19 narrowband emitters from the entire sample (Table~\ref{tab2}). We assume 
$2\sigma$ limiting magnitudes for non-detections, and then classify them if their 
upper limits or lower limits satisfy the colour criteria. Combined with 49 
confirmed members \citep{Shimakawa:2014,Shimakawa:2015b,Shimakawa:2017b}, we adopt 
107 narrowband emitters as the HAE protocluster sample. It should be noted that 
those colour criteria are not guaranteed to select HAEs or to remove foreground 
and background emitters. In particular, part of colour-selected HAEs and rejected 
contaminants are located near boundaries of the criteria, and we cannot securely 
classify them taking account of those photometric errors. Further spectroscopic 
observation is needed to reliably identify whether or not they are protocluster 
members. Lastly, there are 51 unclassified emitters that cannot be classified by 
the colour criteria due to the non-detection in at least one band. These are 
defined as HAE candidates. Given that our target is the dense protocluster region, 
the majority of such faint emitters should be HAEs. The contamination rate of HAE candidates can be roughly expected to be $\sim15$ per cent given 19 rejected 
objects among 126 ($107+19$) brighter emitters.

\begin{table}
\caption{Classification of narrowband emitters in USS1558$-$003. This work employs 
confirmed and colour-selected emitters as the HAE sample, 
which account for 85 per cent of the bright narrowband emitters 
that can be classified by our colour selection criteria. 
We also use HAE candidates in some cases.}
\begin{center}
\begin{tabular}{lrl}
\hline
Class                  & N   & Description \\
\hline
NBEs                   & 177 & NB emitters \citep{Hayashi:2016} \\
HAEs (confirmed)       &  49 & spec-$z$ or \lya\ confirmation \\
HAEs (by colours)      &  58 & selected by r'JKs and Br'Ks \\
other line emitters    &  19 & rejected by r'JKs or Br'Ks    \\
\hline
Total HAEs             & 107 & confirmed + colour-selected \\
HAE candidates         &  51 & faint and ambiguous \\
\hline
\end{tabular}
\end{center}
\label{tab2}
\end{table}%


\section{Results}\label{s3}

Using the constructed HAE database of the protocluster USS~1558, this work compares 
various physical properties of HAEs between higher and lower densities within the 
young protocluster at $z=2.53$. This provides us with insight into the 
environmental dependence on galaxy formation in an early stage of inside-out 
evolution of the galaxy cluster. Our deep \ha\ imaging data enable an investigation 
of the environmental dependence of galaxy characteristics across a diversity of 
substructures in the protocluster on smaller than one physical Mpc (ph-Mpc) scale.


\subsection{Spatial distributions}\label{s3.1}

Spatial distributions of narrowband-selected line emitters can identify structures 
and substructures in and around the forming protoclusters, as demonstrated by the 
past studies 
\citep{Kurk:2004a,Tanaka:2011,Matsuda:2011,Hayashi:2012,Koyama:2013a,Cooke:2014}. 
We caution, however, that \lya\ emitters which can be significantly missing in 
local overdensities \citep{Shimakawa:2017b}. 
Furthermore, our narrowband imaging cannot trace low EW ($<45$ \AA) 
and non \ha-emitting (i.e. passive) galaxies. This should cause errors of density 
estimation which we cannot resolve as it stands. However, we stress that we would 
not miss significantly overdense substructures by HAEs since HAEs are well 
tracing the structure of distant red galaxies ($J-K_s>1.38$; \citealt{Franx:2003})
in this protocluster region \citep{Hayashi:2012}.

To quantify environment, we define a density parameter, the mean projected distance 
$\overline{b}_\mathrm{5th}$, as described in the following manners: 
\begin{equation}
\overline{b}_\mathrm{5th} = 2\sqrt{\frac{1}{\pi\Sigma_\mathrm{Nth}}},
\label{eq5}
\end{equation}
where $N$ is the number of HAEs within a radius of $r_\mathrm{Nth}$ that is the 
distance to the $(N-1)$th neighbour from each HAE. $\Sigma_\mathrm{Nth}$ is the 
number density within the Nth radius ($=N/{\pi r_\mathrm{Nth}^2}$). This paper 
employs $N=5$, and we note that the measured density parameters maintain relative 
consistencies even if we choose different $N$ values in the single-digit range. 
$\overline{b}_\mathrm{5th}$ expresses how close galaxies are to each other in 
projected space at $z\approx2.53$ in the sense of lower values indicating higher 
densities. Also, this value can be regard as the familiar $\Sigma_\mathrm{5th}$ 
as seen in the equation \ref{eq5}. The major reason to use 
$\overline{b}_\mathrm{5th}$ is that the mean projected distance can be used as a 
mean impact parameter which has been commonly employed in the \lya\ absorption 
analyses (e.g. \citealt{Wolfe:2005,Fumagalli:2010,Rakic:2012}). 
Since our forthcoming paper will discuss environmental effects on \lya\ photon 
escape fraction as preliminary reported in \citet{Shimakawa:2017b}, the mean 
projected distance is an ideal indicator to quantify the local overdensities. 

\begin{figure}
\centering
\includegraphics[width=.95\columnwidth]{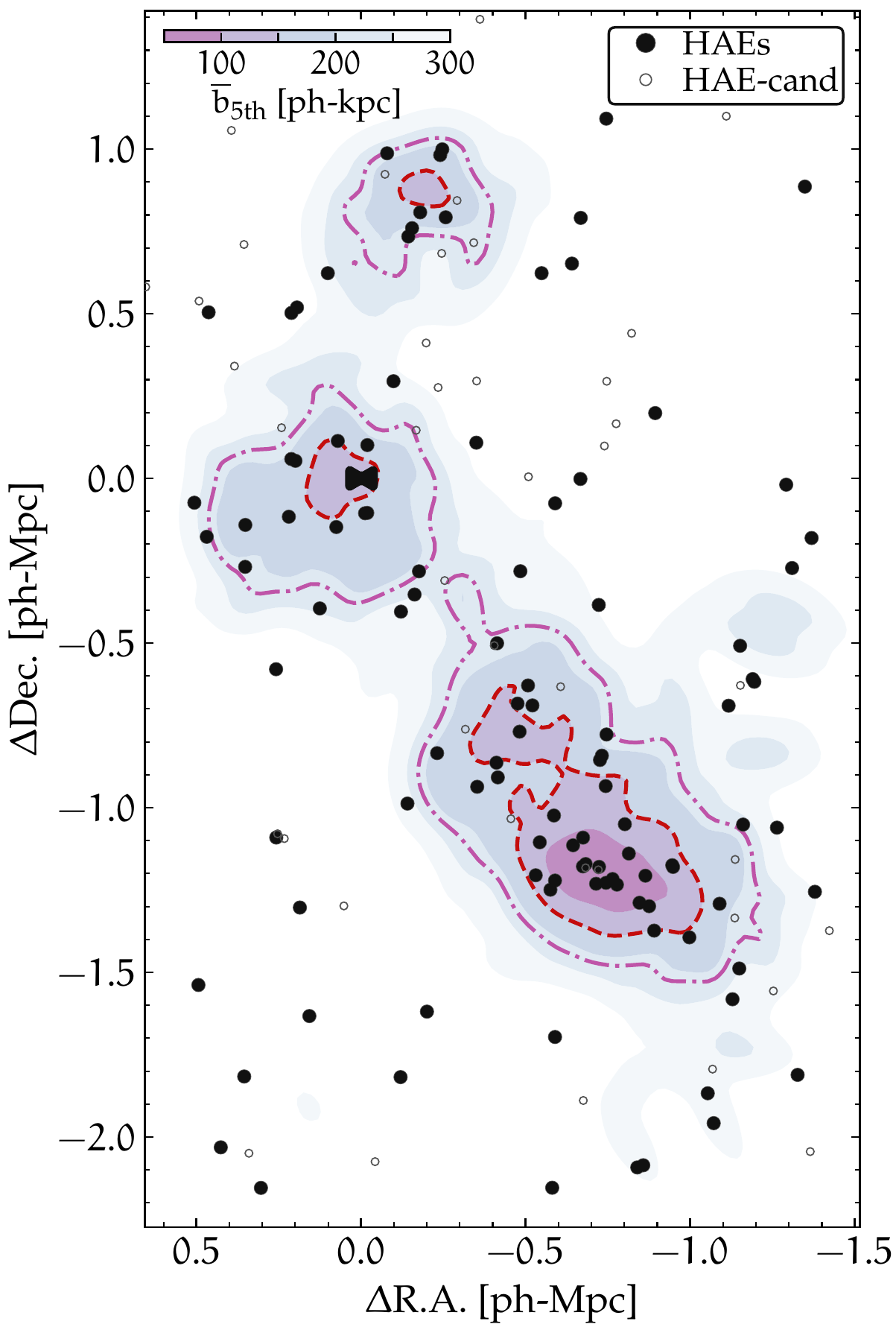}
\caption{Spatial distribution of HAEs (filled circles) and HAE candidates (open 
circles). The bow-tie is the radio galaxy, USS~1558$-$003. Filled contours show 
the 5th mean projected separations of $\overline{b}_\mathrm{5th}<100$, 100--150, 
150--200, 200--250, 250--300 ph-kpc, respectively. The contour map is smoothed by 
a 2D Gaussian kernel with $\sigma=1.5$ arcmin. The purple dash-dot contour line 
divides the sample into group and intergroup regions 
($\overline{b}_\mathrm{5th}=220$ ph-kpc or 50th percentile line). We also identify 
them at the top quartile high-density region as supergroup HAEs with the mean 
projected separation lower than $\overline{b}_\mathrm{5th}=150$ ph-kpc as marked 
by the red dashed curve.}
\label{fig3}
\end{figure}

The spatial distribution and density structure of HAEs in USS~1558 are given by 
Figure~\ref{fig3}. As reported by \citet{Hayashi:2016}, our MDC campaign succeeds 
in increasing the protocluster sample by 1.6 times as compared to the past survey 
\citep{Hayashi:2012}. In our sample, 39 out of 107 HAEs are newly discovered as 
protocluster members and we also have additional 51 candidates (Table~\ref{tab2}). 
A typical mean projected separation of $\overline{b}_\mathrm{5th}=220$ ph-kpc 
indicates that our target is associated with remarkably high overdensities, roughly 
corresponding to 15--20 higher number density than the general field. It reaches to 
$\overline{b}_\mathrm{5th}<150$ ph-kpc in the top quartile of high-density regions, 
which is comparable in size to virial radii of typical HAEs with a few times 
$10^{12}$ \msun\ halo masses \citep{Cochrane:2017}. 
Such a heavy congestion implies that those haloes would overlap with each other 
sometimes physically and more easily along the line of sight, and also they might be 
sharing large common haloes with their neighbours. 

To examine the effect of dependence within the protocluster, we divide the entire 
HAE sample into the {\it group} and {\it intergroup} samples by the median 
$\overline{b}_\mathrm{5th}$ value of 220 ph-kpc (dash-dot curve in Fig.~\ref{fig3}). 
This criterion also corresponds to the $1\sigma$ local overdensity in the survey 
area. Furthermore, we define a {\it supergroup} HAE sample to investigate 
environmental effects on some physical properties in the extraordinarily high 
density regions that reach $\overline{b}_\mathrm{5th}<150$ ph-kpc, which corresponds 
to the top quartile of overdensities in the protocluster (dashed curve in 
Fig.~\ref{fig3}). 
While our deep data should enable the finest structure mapping for such a high-$z$ 
protocluster ever, one should note that the classification of the sample is 
data-driven rather than scientifically justified. We have no clear answer yet to 
properly evaluate substructures in high-$z$ protocluster for such \ha-selected 
sample in the restricted area. Thus, this work simply compares galaxy properties 
between higher and lower densities in the same protoclustrer region, and leaves 
this issue for future study. \hi\ tomographic mapping by deep far-UV spectroscopic 
analyses \citep{Lee:2014a,Lee:2016} could be another, more direct solution to 
identify the substructures, although it is very difficult with current facilities 
to reach the spatial resolution of $\sim1$ comoving Mpc (hereafter co-Mpc) necessary 
to resolve the protocluster halo by this kind of survey.


\subsection{Star-forming main sequence}\label{s3.2}

This section firstly repeats the derivation of stellar masses and SFRs previously 
done by \citet{Hayashi:2016}, but for the updated HAE samples. We then compare those 
between the group and intergroup environments. This work revises previous estimates 
of these parameters as described in the following sentences, implying slight 
systematic differences from the previous research. However, we stress that these 
revisions do not affect our conclusions. 

We obtained the narrowband line flux ($F_\mathrm{NB}$) and rest EW by the following 
formula, 
\begin{eqnarray}
F_{NB} &=& \Delta_{NB} (f_{NB}-f_{Ks'}) \label{eq6} \\
\mathrm{EW}_{NB,rest} &=& \frac{F_{NB}}{f_{Ks'}}(1+z)^{-1} \label{eq7}
\end{eqnarray}
where $\Delta_{NB}$ is full-width-half-maximum (FWHM) of $NB2315$ filter 
(260 \AA). $f_{Ks'}$ is the $K_s$-band flux density including the fixed colour term 
correction (0.07 mag) derived from model-inferred SEDs of HAEs (Fig.~\ref{fig14}). 
We checked the impacts of this assumption by comparing measured line fluxes with 
those including the colour term correction in which we employ SED-inferred continua 
at the narrowband wavelength instead of $f_{Ks'}$. We find that the effect of the 
colour term variation on individual line fluxes should be negligible because they 
are consistent with each other within the standard dispersion of 0.065 dex. 
Indeed, our HAEs have high EW$_\mathrm{H\alpha}$ (the median value 
of 180 \AA\ in the rest frame), and thus, line flux is dominant at the narrowband 
filter wavelength and the colour-term variation does not have a large effect such 
that similar studies caution for line emitters at lower redshifts (e.g. 
\citealt{Vilella:2015}).
Thus, we stress that such a small uncertainty does not affect our results and 
conclusions. 

After that, we carry out SED-fitting based on the SED-fitting code {\sc fast} 
distributed by \citet{Kriek:2009}. 
We used $B$, $r'$, $z'$, $J$, $H$, $F160W$, $K_s$ band photometry 
for the SED fitting. We should note that {\sc fast} code does not include nebular 
emission line components. This may cause systematic errors of stellar mass 
measurements of our sample, especially for lower-mass HAEs. \citet{Atek:2011} 
reported that SED-inferred stellar masses for high EW ($>200$ \AA) galaxies without 
removing emission lines tend to be overestimated by a factor of two on average (see 
also e.g. \citealt{Wuyts:2007,Schaerer:2013,Stark:2013,Salmon:2015}). 
We believe that such a systematic error should be much smaller in our case since 
our deep $F160W$ ($5\sigma=27.5$ mag) and $K_s$ ($5\sigma=24.5$ mag) data that are 
critical for stellar mass estimations, are not affected by strong \ha, \oiii, \hb\ 
emission lines.
SED fitting was conducted assuming a fixed 
redshift of $z=2.53$, based on the stellar population model of \citet{Bruzual:2003}, 
the \citet{Calzetti:2000} extinction curve, the \citet{Chabrier:2003} IMF, and fixed 
metal abundance of $Z=0.004$ ($0.2 Z_\odot$). We assume delayed exponentially 
declining star formation histories (SFR $\propto~t~exp(-t/\tau)$) where $\tau$ is 
allowed to be between $10^9$ and $10^{11}$ yr. Population ages between 10$^{7.6}$ 
and 10$^{9.4}$, and stellar extinction ($A_V$) between 0 and 2.4 mag are given, 
respectively. These values do not significantly affect stellar mass estimations, 
but the measurements of dust extinction systematically depend on input parameters. 
For example, measured dust extinctions ($A_V$) in 80 per cent of HAEs change within 
a range of $\pm20$ \% uncertainties (and residuals vary even more) if we apply free 
parameters for $\tau$ and age given by the entire library. 
We ignore such a model dependency through the paper because this work solely focuses 
on the relative comparison of HAE samples selected in the same way, and based on the 
same datasets. $1\sigma$ errors of obtained physical parameters are estimated by 100 
Monte Carlo simulations attached with the {\sc fast} code. 

\begin{figure}
\centering
\includegraphics[width=0.9\columnwidth]{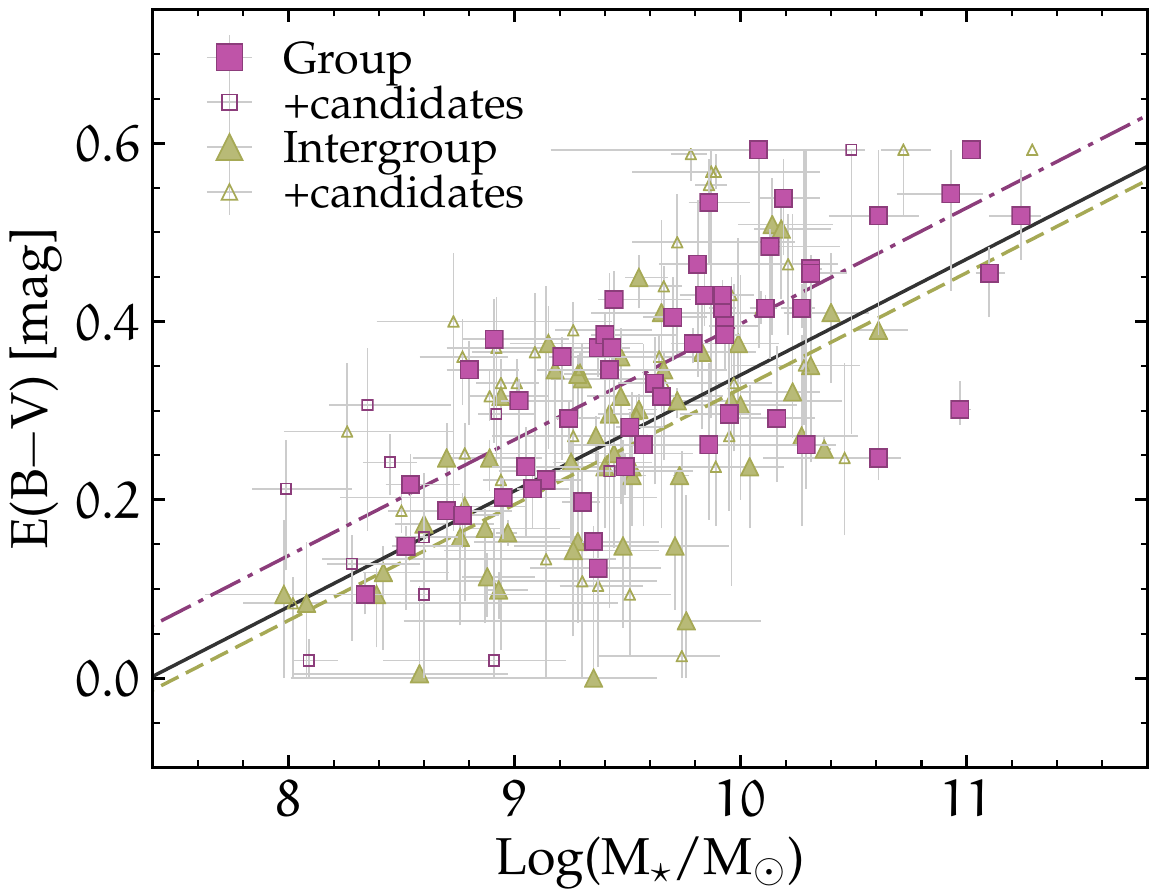}
\caption{Stellar mass versus dust reddening, E(B$-$V) for HAEs (filled) and HAE 
(open) candidates. Purple squares and yellow triangles show the HAE samples in the 
group and intergroup regions, respectively. Solid, dash-dot, and dashed lines 
indicate the best fit line for the entire HAEs, group HAEs, and intergroup HAEs, 
respectively, where the latter two regressions assume the same slope as that for the 
entire HAEs. Errorbars are $1\sigma$ errors based on the {\sc fast} measurements.}
\label{fig4}
\end{figure}

The derived stellar masses and dust reddening are presented in Figure~\ref{fig4}. 
The data are divided into group and intergroup samples by colour. As many of 
researchers have previously reported 
(e.g. 
\citealt{Garn:2010,Reddy:2010,Sobral:2012,Ibar:2013,Zahid:2013b,Whitaker:2014b,Koyama:2015}), 
these two parameters correlate with each other. The environmental dependence of 
measured stellar masses and dust extinctions of HAEs is discussed in \S\ref{s3.4} 
and \S\ref{s3.5}. 

We estimate SFR$_\mathrm{H\alpha}$ of HAEs from the \ha\ narrowband fluxes based on 
the \citet{Kennicutt:1998} prescription assuming the \citet{Chabrier:2003} IMF (a 
factor of 1.7 reduction in SFR from the standard \citealt{Kennicutt:1998} 
calibration). We correct for \nii$\lambda\lambda$6550,6585 line contamination to the 
narrowband flux by using the stellar mass--\nii/\ha\ relation for star-forming 
galaxies at a similar redshift (equation 20 in \citealt{Steidel:2014}). 
Although \citet{Steidel:2014} have selected the sample based on UV, 
sampling effect would be negligible since their mass--metallicity relation is 
consistent with that of H-band selected galaxies at similar redshift 
\citep{Sanders:2015}. 
In light of the small scatter in the mass-metallicity relation (0.1 dex), the 
typical error in \nii\ flux can be estimated to be $<12$ per cent. We also calculate 
a dust extinction correction for the \ha\ emission line based on the SED-inferred 
stellar extinction ($A_V$) by {\sc fast} and the \citet{Calzetti:2000} extinction 
law under the assumption that 
E(B$-$V)$_\mathrm{stellar}$ = E(B$-$V)$_\mathrm{nebular}$. 
Although we have to rely on this assumption because we have no \hb\ line data for 
the Balmer decrement, this hypothesis is relatively reasonable at high redshifts 
\citep{Reddy:2015}. One should note that ratio of nebular extinction to stellar 
extinction likely depends on galaxy properties, especially on SFR 
\citep{Price:2014,Reddy:2015}. This may mean that our SFR measurement may 
underestimate mainly the SFR of active and dusty star-forming objects which are 
expected at the massive end. However, it is challenging to estimate proper SFR 
values of such dusty starbursts in the rest optical regime in any case 
\citep{Dannerbauer:2014,Tadaki:2017}. More importantly, such massive and/or dusty 
HAEs are preferentially located in higher densities (\S\ref{s3.4} and \ref{s3.5}), 
and thus, our comparison results and conclusions (\S\ref{s5}) will be enhanced once 
we could correct those nebular extinctions more precisely. Measured parameters of 
individual HAEs are summarised in Appendix \ref{a1} (Table~\ref{tab5}).

\begin{figure*}
\centering
\includegraphics[width=0.9\textwidth]{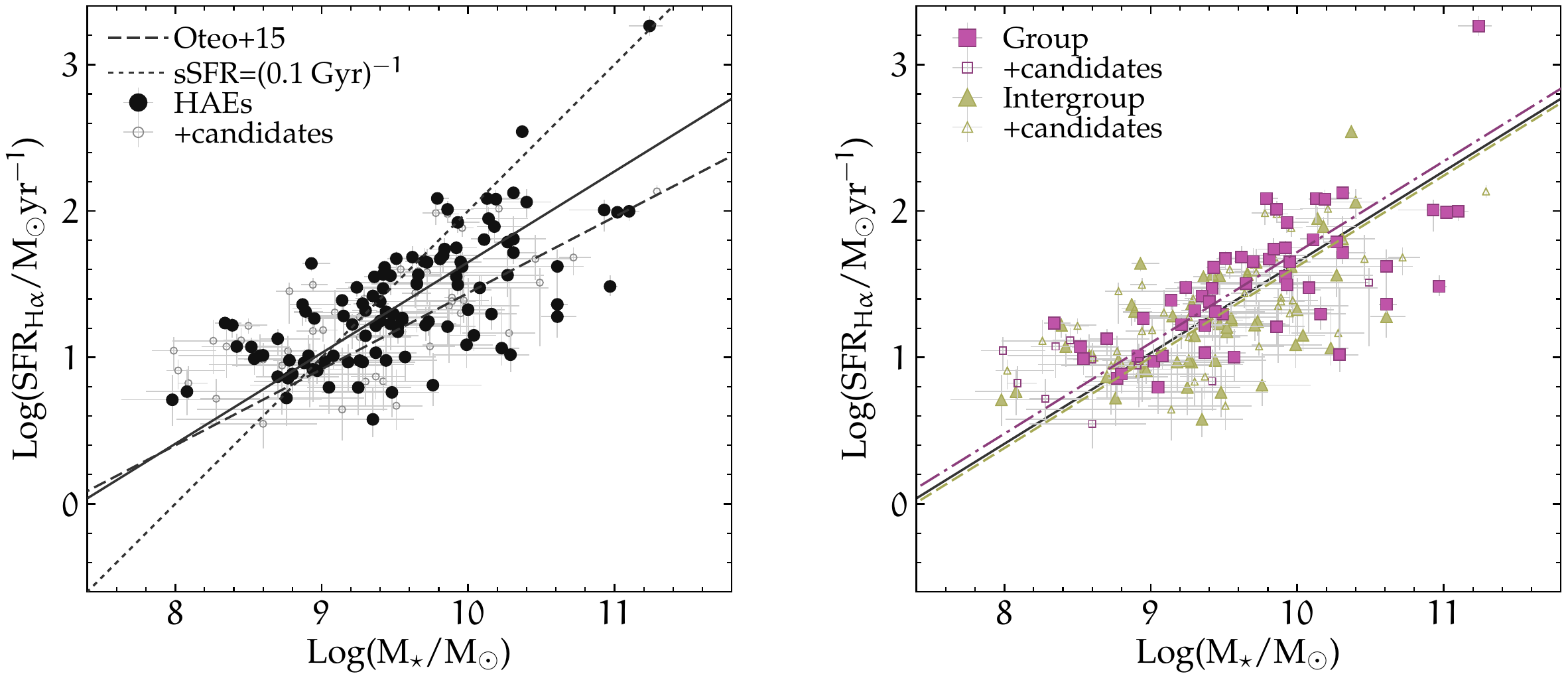}
\caption{Stellar mass versus SFR from dust-corrected \ha\ luminosity. {\it Left} 
and {\it right} panel show the same results, but the right figure changes the colours 
of HAEs in group (purple squares) or intergroup (yellow triangles) region. Filled and 
open symbols indicate HAEs and HAE candidates (Table~\ref{tab2}), respectively. 
The black solid and dashed line show M$_\star$--SFR relations for our HAE sample and 
HAEs in the general field at $z=2.2$ \citep{Oteo:2015}, respectively. 
\citet{Oteo:2015} have corrected dust extinction with a mass-dependent prescription 
\citep{Garn:2010}. Purple dash-dot and yellow dashed lines are the median-fitted 
lines for group and intergroup HAEs, assuming the same slope 
(SFR $\propto$ M$_\star^{0.62}$) as that for the entire HAEs. The black dotted line 
indicates sSFR = (0.1 Gyr)$^{-1}$.}
\label{fig5}
\end{figure*}

Figure~\ref{fig5} shows the dust-corrected SFRs of the HAEs as a function of stellar 
mass. Our HAE sample establishes a mass--SFR relation known as the star-forming 
main sequence, which is discovered by \citet{Brinchmann:2004} in the local Universe 
and also by \citet{Daddi:2007a,Elbaz:2007,Noeske:2007} at high redshifts, and then 
followed and improved by so many studies (e.g. 
\citealt{Whitaker:2012,Speagle:2014,Oteo:2015,Shivaei:2015,Tomczak:2016}). The SFR 
limit of our narrowband observations is 2.4 or 6.0 \msun~yr$^{-1}$ with 
$A_\mathrm{H\alpha}=$ 0 or 1 mag, respectively. Because of this SFR limit, our 
sample is biased toward highly active HAEs in the low-mass regime especially below 
M$_\star\lesssim10^{9.4}$ \msun. In other words, our HAE sample is relatively 
complete above this stellar mass range, although low line EW ($<45$ \AA) galaxies 
and heavily-obscured HAEs will be absent. 

Figure~\ref{fig5} shows that our HAE sample well follows the star-forming main 
sequence reported by \citet{Oteo:2015} for \ha-selected galaxies in the general 
field at $z=2.2$. The typical offset value from the main sequence of 
\citet{Oteo:2015} is estimated to be 0.15 dex. This small offset may be caused by 
redshift evolution ($z=2.2$ vs. $z=2.5$), environmentally-driven effects, or 
different extinction corrections. Besides, our deep \ha\ imaging goes down to by 
2.4 times fainter \ha\ luminosities than the previous data \citep{Hayashi:2012}, 
which enables discovery of 20 active low-mass HAEs (M$_\star<10^9$ \msun) with 
very high specific SFRs (sSFRs) $>$ (0.1 Gyr)$^{-1}$. This paper does not touch 
the detail of such unique emitters since those are already discussed by 
\citet{Hayashi:2016}. 

The right panel in the Fig.~\ref{fig5} represents the same as the left, but here we 
separate the sample into HAEs in group ($\overline{b}_\mathrm{5th}<220$ ph-kpc) and 
intergroup ($\overline{b}_\mathrm{5th}>220$ ph-kpc) environments as classified in 
the previous section. The median SFR of HAEs in group and intergroup regions are 
30 \msun yr$^{-1}$ and 17 \msun yr$^{-1}$, respectively, and the Kolmogorov--Smirnov 
(KS) test suggests the dense group environments tend to have more active HAEs by the 
$p$-value of 0.005. For a given stellar mass, however, those two populations are no 
longer statistically different ($p=0.353$) as inferred from the comparison of the 
offset from the median-fitted main sequence (hereafter $\Delta\mathrm{MS}$) in the 
stellar mass range of $10^{9-10.5}$ \msun. This means that HAEs shift upper 
rightward approximately along the main-sequence depending on the local overdensity. 
Such trends broadly agree with \citet{Koyama:2013a,Koyama:2013b} that have reported 
as up-rightward shift along the main sequence in the protocluster at $z=2.2$ as 
compared to the field at the same redshift. The section \S\ref{s3.5} summarises the 
results of the KS test for physical parameters. 

\begin{figure*}
\centering
\includegraphics[width=0.6\textwidth]{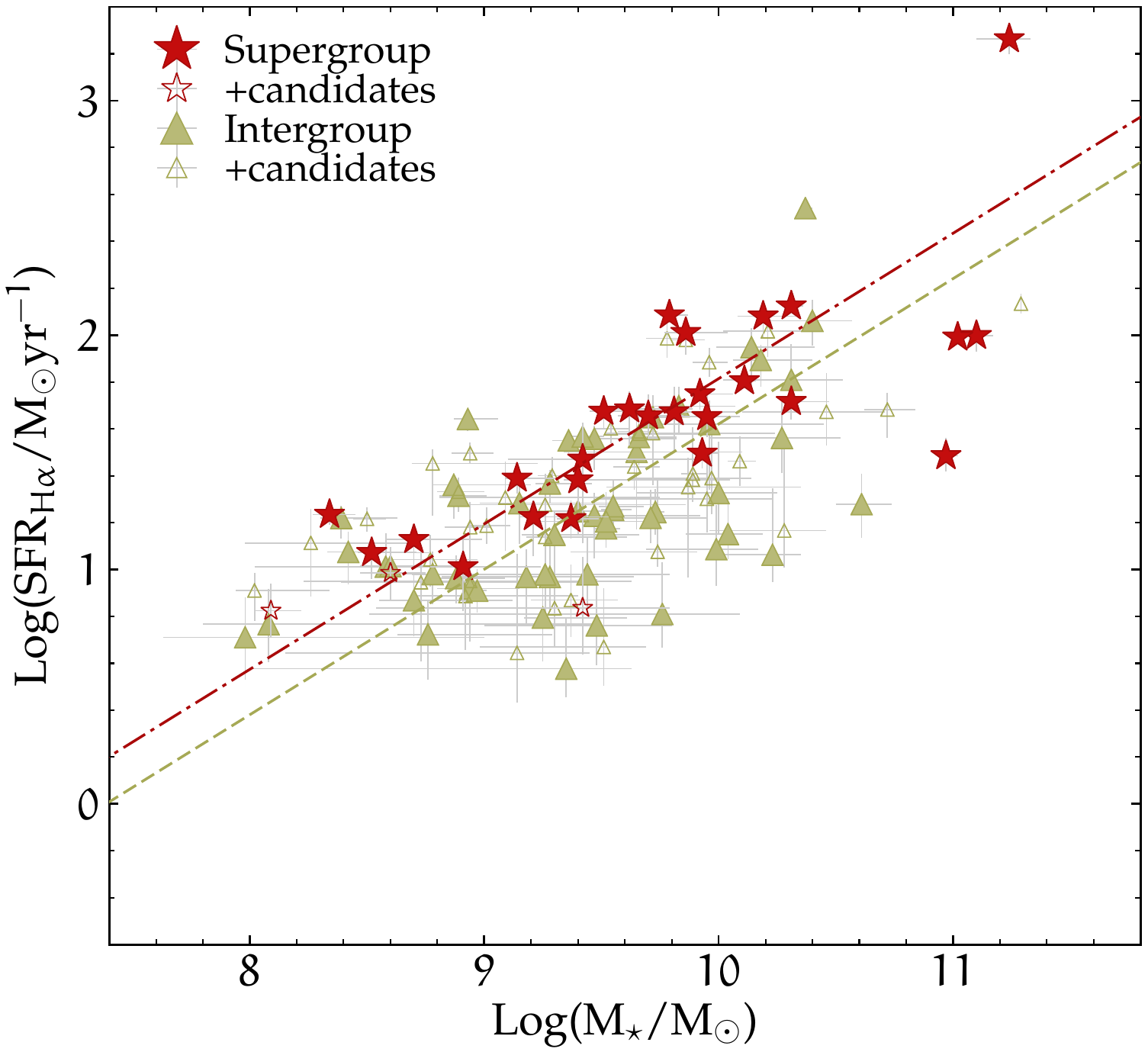}
\caption{The same as Fig.~\ref{fig5}, but here we compare HAEs in between the 
supergroup (top quartile of overdensities) shown by red star symbols and the 
intergroup regions. Red dash-dot and dashed lines are the median-fitted lines 
for supergroup and intergroup HAEs assuming the same slope as the entire HAEs, 
respectively.}
\label{fig6}
\end{figure*}

Surprisingly, however, we identify even clearer SFR differences across local 
densities when we compare those between the supergroup and the intergroup 
(Fig.~\ref{fig6}). The KS test indicates those two populations are discriminable 
from each other ($p=0.03$) even if we compare those for a given stellar mass i.e. 
$\Delta\mathrm{MS}$. 81\% of the supergroup HAEs consist of the densest star-forming 
clump in SW area in the protocluster (Fig.~\ref{fig3}) where the mean projected 
separation reaches  $\overline{b}_\mathrm{5th}<150$ ph-kpc or $<640$ ph-kpc in the 
3-D space. In the supergroup region, HAEs with M$_\star<10^{10.4}$ \msun\ are 
located at the upper envelope of the main sequence of HAEs in the intergroup 
regions, suggesting that extremely high densities accelerate galaxy formation. 
Seven out of 26 HAEs in the supergroup region have projected distances of $<30$ 
ph-kpc to the nearest neighbours, and thus, some of those may be merger-induced 
starburst galaxies. 

One critical concern for clear SFR enhancement in the supergroup regions is 
the uncertainties of extinction corrections for individual HAEs. However, we find 
a significant excess even without dust corrections ($p=0.03$). Another issue is a 
systematic error due to the filter response function of the $NB2315$ narrowband 
filter (that is not a perfect top-hat function). We run another simulation to test 
this possibility, and then find that a quite drastic situation should be required 
to explain the difference only by the error of filter flux loss (see Appendix 
\ref{a3.2}). Thus, enhanced star formation of HAEs in the extremely dense group 
regions is more likely an intrinsic phenomenon. That said, deep follow-up NIR 
spectroscopy is highly desired to confirm SFR enhancement in the supergroup regions. 

The standard deviation of the main sequence is also worth investigating 
\citep{Lin:2012,Koyama:2014}, for instance, to test if dense environments enhance 
the merger rate and/or then increase the number of starburst galaxies 
\citep{Gottlober:2001,Genel:2014}, and drive environmental quenching 
\citep{Wetzel:2013,Gobat:2015}. However, the observed scatter of the star-forming 
main sequence show consistent values of 0.33 and 0.34 dex between group and 
intergroup regions, which also agree with typical value of $\sim0.3$ dex in the 
general field at the similar redshifts reported by past studies (e.g. 
\citealt{Speagle:2014,Shivaei:2015}). More interestingly, HAEs at 
M$_\star<10^{10.4}$ \msun\ in the densest regions ($\overline{b}_\mathrm{5th}<150$ 
ph-kpc) have a tighter main sequence ($\sigma=0.19$ dex or 1.5 times smaller than 
that of intergroup HAEs on a linear scale). We should note, however, that our \ha\ 
survey may miss dust-hidden starbursts and we may underestimate dusty HAEs in our 
sample. Also, our narrowband survey cannot detect objects with declining star 
formation with low \ha\ EWs. These uncertainties may lead us to underestimate the 
main sequence scatter. 

\subsection{Mass--size distribution}\label{s3.3}

\begin{figure*}
\centering
\includegraphics[width=0.7\textwidth]{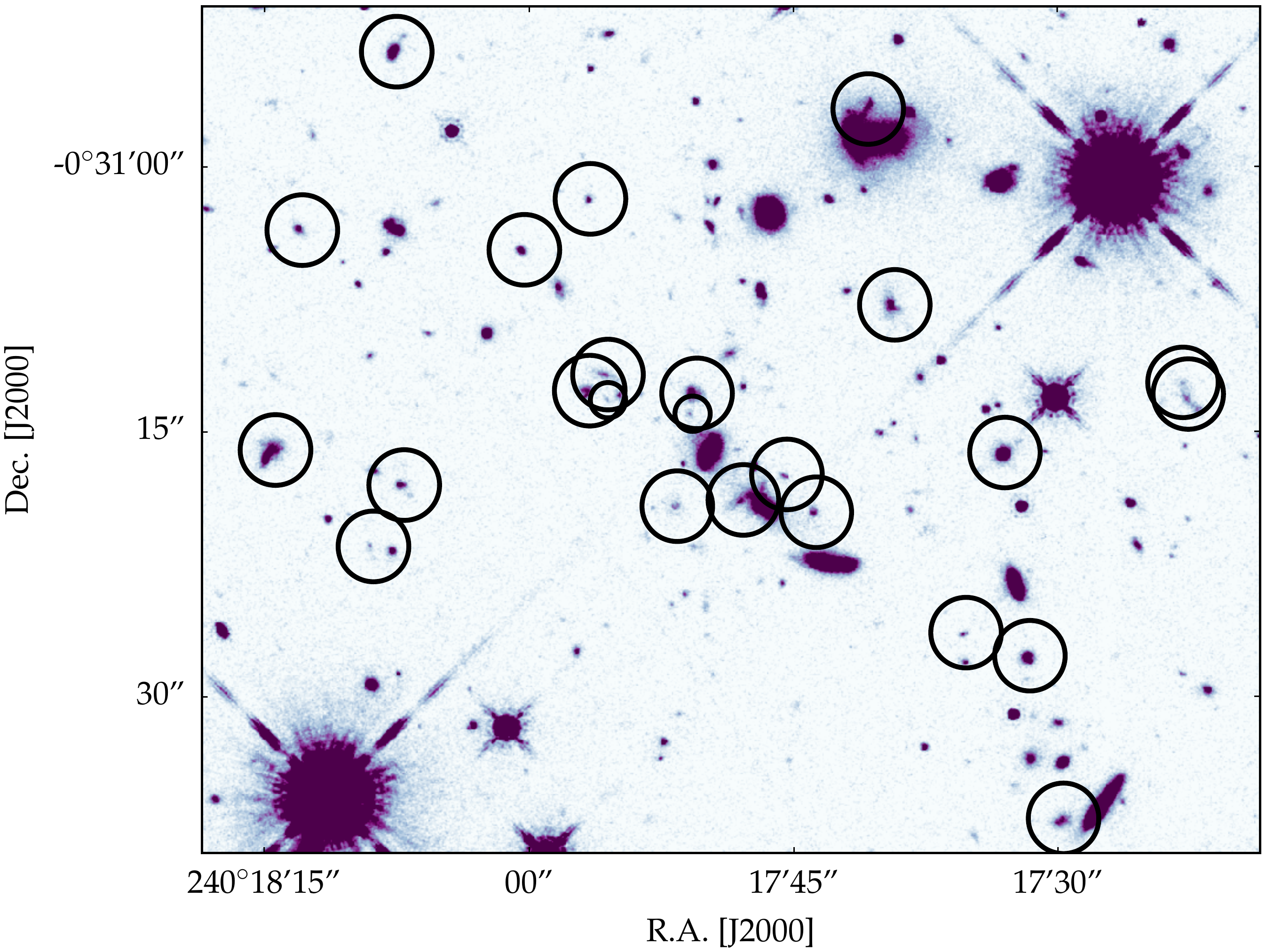}
\caption{F160W image form WFC3/HST ($1.0\times0.8$ arcmin$^2$) around the 
peak density in the USS~1558 protocluster. Large and small circles indicate 
HAEs and HAE candidates selected by this work, respectively.}
\label{fig7}
\end{figure*}

Size estimation of HAEs in the protocluster and comparison to stellar sizes across 
environments allow an investigation of the environmental effects on size growth 
histories at high redshifts. For example, overdense environments may involve 
the higher chance of galaxy mergers \citep{Okamoto:2000,Gottlober:2001,Hine:2016}. 
Such merger events will change the size growth rate as a function of total stellar 
mass. Specifically, it is expected that major mergers and minor mergers increase the 
effective radii $R_e$ of stellar components scaled as 
$\delta\log$($R_e$)/$\delta\log$(M$_\star$) $\sim1$ and 2, respectively 
\citep{Bezanson:2009,Naab:2009,Dokkum:2010}. Mergers would also drive the gas 
towards the centre and may induce more centrally concentrated star formation 
(e.g. \citealt{Mihos:1996,Teyssier:2010,Bournaud:2015}). So 
far, many studies have investigated the environmental dependence of mass and size 
evolution in high-$z$ clusters 
\citep{Papovich:2012,Lani:2013,Gobat:2013,Newman:2014b,Belli:2014,Tran:2017,Kubo:2017}, 
however, the conclusion is still controversial, especially in rich systems of 
star-forming galaxies like USS~1558. The existing F160W image in USS~1558 taken by 
HST (Fig.~\ref{fig7}) allows us to derive the sizes of rest-frame optical light for 
the HAEs, based on the model profile fitting technique by {\sc galfit} 
\citep{Peng.CY:2010}. 

\begin{figure*}
\centering
\includegraphics[width=.9\textwidth]{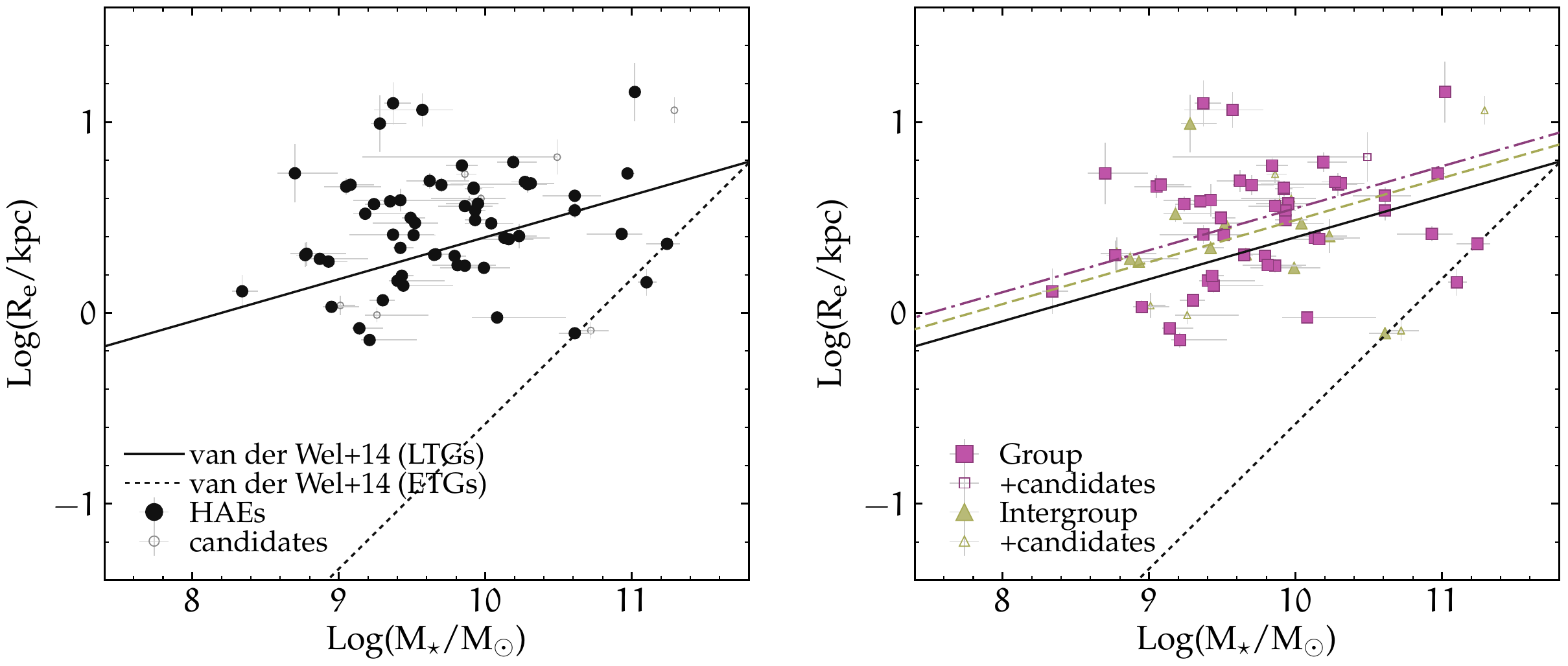}
\caption{Stellar mass versus effective radius. Two panels show the same results, 
but symbols are changed by different environments in the right figure. The symbols 
are the same as shown in Fig.~\ref{fig5}. Dashed and dash-dot lines indicate 
mass-size relation of star-forming galaxies and quiescent galaxies reported by 
\citet{Wel:2014}. In the right panel, purple dash-dot and yellow dashed lines show 
the median-fitted lines under the same slope of the black solid line. PSF objects 
are estimated to be $\log(\mathrm{R_e}/kpc)\sim-0.7$ dex.}
\label{fig8}
\end{figure*}

We estimate effective radii $R_e$ of HAEs from the F160W image by using the 
{\sc galfit} version 3.0 code \citep{Peng.CY:2010}. The F160W filter does not include 
strong \hb\ and \oiii$\lambda\lambda4960,5008$ emission lines, and thus, the derived 
$R_e$ should be tracing only stellar components of the target HAEs. Since our HST 
observation mainly focused on the overdense regions \citep{Hayashi:2016}, only 47 (1) 
and 12 (7) bright HAEs (or HAE candidates) with F160W $<25$ mag (corresponds to 
$\gtrsim10\sigma$ detections) are available. Initial parameters for the model fitting 
(i.e. central position, total magnitude, half light radius, axis ratio, and position 
angle) incorporate the results from SExtractor \citep{Bertin:1996}. We initially set 
a Sersic index of 1.5, and allow it to be 0.2--8 in the fitting. We 
employed the PSF model from the 3D-HST \citep{Skelton:2014} that has better 
signal-to-noise ratio than that of the PSF model made by the composite image of 
bright stars in our survey field. In addition, we correct the wavelength dependence 
of size estimation caused by radial colour gradients of galaxies that largely depend 
on stellar mass \citep{Wuyts:2012,Wel:2014} according as the empirical relation 
provided by \citet{Wel:2014}, which converts the obtained effective radii ($R_e$) to 
that at the rest frame wavelength of 5000 \AA. This empirical conversion is 
applicable only for star-forming galaxies, which can be eligible for most HAEs. One 
should note that the correction factors are at most 4 per cent for our sample, which 
are negligibly small. Our size measurement includes photometric errors coupled with 
the fitting errors from {\sc galfit}, but does not consider further possible 
uncertainties as noted by \citet{Wel:2012}. Therefore, our measured $R_e$ may have 
additional errors of at most 0.3 dex depending on these actual surface brightness 
and radial profiles \citep{Wel:2012}. 

The obtained results are shown in Fig.~\ref{fig8} and summarised in 
Table~\ref{tab5}. Unfortunately, the sample size in the intergroup regions is 
obviously insufficient, and thus, we can only make a poor comparison between two 
populations of different densities. We also estimate offsets from the mass--size 
relation \citep{Shen:2003} of star-forming galaxies at $z=$ 2--2.5 reported by 
\citet{Wel:2014}, which is defined by $\Delta{Re_{V14}}$. We obtained those typical 
values of $\Delta{Re_{V14}}=0.15$ and 0.08 dex in group HAEs and intergroup HAEs. 
We cannot identify the apparent deviation of size distributions in our samples, 
since those still fall within $1\sigma$ scatter ($\sim0.19$ dex) of 
\citet{Wel:2014}, respectively. 
We also confirm that the stellar size distribution of our HAEs is 
consistent with that of HAEs in general field at $z=2.2$ \citep{Stott:2013b}.
According to the KS test, we cannot rule out that 
groups of HAEs belong to the same parent sample as those in the intergroup 
regions along the mass--size relation ($p=0.379$, see \S\ref{s3.5}). 
In fact, rather, it is hard to judge how our HAEs differ in 
mass--size distribution within our sample and galaxies in the previous studies 
due to restricted sample size. Furthermore, our small sample cannot constrain any 
relation between stellar mass and effective radius with the high $p$-value of 
0.15 by the Spearman's rank correlation coefficient. Considering these factors, 
our data should not be enough to discuss size evolution and we decide not to delve 
into the size distribution in this paper. 

On the other hand, we identify two and one compact HAEs in the higher and lower 
densities in the protocluster, respectively, that follow the mass--size relation of 
early type galaxies at the similar redshift slice \citep{Wel:2014}. It would be 
intriguing to carry out follow-up spatially-resolved analyses of those star-forming 
activities based on integral field spectroscopy of \ha\ line or dust continuum to 
understand the formation mechanisms of passive red sequence objects in local galaxy 
clusters.


\subsection{H$\alpha$ luminosity and stellar mass functions}\label{s3.4}

It is also essential to compare number distributions of such physical properties 
across overdensities, since we can expect that those distribution functions would 
depend on surrounding environments inferred from the substantial excess of very 
massive quiescent galaxies in the cluster centre in the local Universe. Indeed, 
previous studies have shown the enhancement of massive star-forming galaxies in 
protoclusters as compared to the general fields \citep{Steidel:2005,Koyama:2013b}. 
Our extended HAE sample allows us to quantify and advance such past achievements by 
determining  number distributions as a function of \ha\ luminosity and stellar mass 
based on the \citet{Schechter:1976} function. 

The key issue in a derivation of distribution functions is the completeness 
correction (e.g. \citealt{Sobral:2009,Sobral:2013,Sobral:2014}). We correct the 
numbers of the HAE sample based on their narrowband magnitudes as described in 
Appendix \ref{a2}, which includes the correction of both detection and selection 
completeness derived by the Monte Carlo simulation. Our procedure is similar to the 
technique by \citet{Sobral:2009,Sobral:2013} that have conducted completeness 
correction in \ha-selected galaxies by narrowband imaging surveys. 

However, we should still miss a bunch of HAEs due to the incompleteness of the 
following colour--colour selections that were conducted to remove other line 
contaminants (\S\ref{s2.3}). This additional colour selection produced 51 
unclassified narrowband emitters defined as HAE candidates (Table~\ref{tab2}), 
which are a lack of photometry of at least one band among $B, r', J, Ks$ filter 
bands. Since 86 per cents of those are low-mass objects with stellar masses lower 
than $10^{10}$ \msun, the colour--colour selection processes particularly miss 
the objects at the faint end. To minimise this impact, we compromise by adding a 
correction of 15 per cent contaminations for the number of HAE candidates 
(\S\ref{s2.3}). We then use the combined sample of HAEs and HAE candidates for the 
fitting of the Schechter function. This control sample should provide us with a 
relatively-complete HAE sample that is less affected by the colour--colour 
selection criteria, and enable fairer results for our analyses. One should note 
that our correction process cannot recover HAEs with EW$_\mathrm{H\alpha}$ lower 
than 45 \AA\ in the rest frame.

\begin{figure*}
\centering
\includegraphics[width=1.0\textwidth]{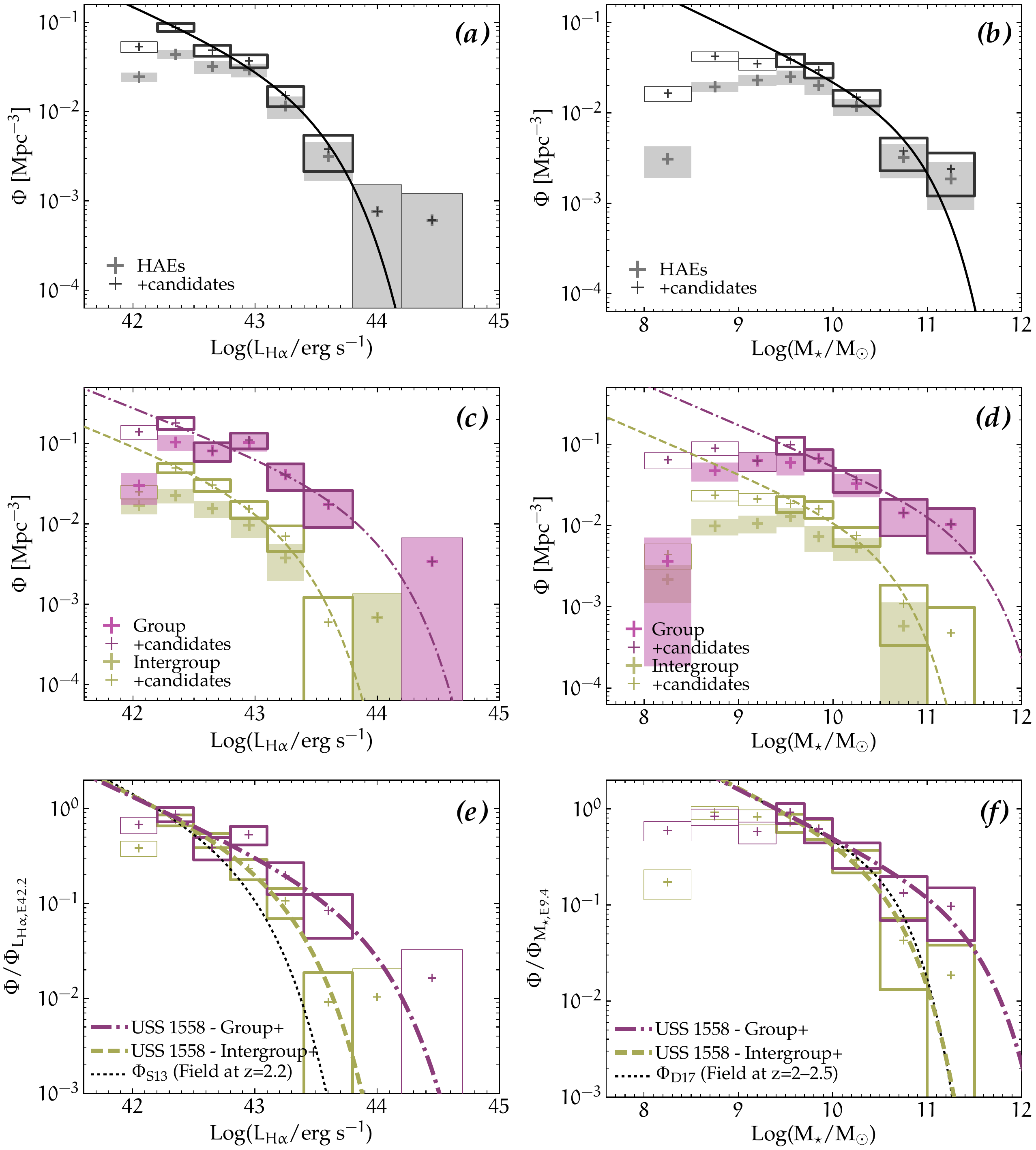}
\caption{\ha\ luminosity ({\it left}) and stellar mass ({\it right}) 
distributions of our HAE samples. Filled boxes indicate the HAE sample and open 
ones include HAE candidates with 15 per cent contamination correction. (a,b) 
show the entire protocluster region, and (c,d,e,f) represent the dense groups 
(purples) and the intergroup regions (yellows). In the panel (e,f), the \ha\ 
luminosity and stellar mass functions are normalised by the values at 
$L_\mathrm{H\alpha}=10^{42.2}$ erg~s$^{-1}$ and M$_\star=10^{9.4}$ \msun, 
respectively. Black solid, purple dash-dot, and yellow dashed curves are 
best-fitted Schechter functions for the protocluster, group, and intergroup 
samples, respectively. The fitting results and the scaling factors are 
listed in Table~\ref{tab3}. Black dotted curves in (e) and (f) indicate the 
distribution functions of HAEs at $z=2.2$ in the random fields by 
\citet{Sobral:2013} and star-forming galaxies at $z=$ 2--2.5 in the random 
fields by \citet{Davidzon:2017}, respectively.}
\label{fig9}
\end{figure*}

\begin{table*}
\centering
\caption{Results of Schechter function fitting for \ha\ luminosity and stellar 
mass bins. The first column specifies the sample. The second and fifth columns 
show the characteristic \ha\ luminosity and stellar mass, respectively. The third 
and sixth columns represent normalisations for \ha\ luminosity and stellar mass, 
respectively. Our line fitting assumes the fixed slope $\alpha=-1.6$ and $-1.5$ 
for \ha\ luminosity and stellar mass, respectively. The fourth and seventh columns 
show the scaling factor used when we normalise and then compare those with each 
other (bottom panels in Fig.~\ref{fig9}). The stellar mass function for the group 
HAE sample can be also fitted by the following power-law function, 
$\log(\Phi\mathrm{_{M_\star}/{Mpc}^{-3}})=3.696\pm0.063-0.5\times\log$(M$_\star$/M$_\odot$).}
\begin{tabular}{lcccccc}
\hline
Sample & log($L_\mathrm{H\alpha}^\ast$/erg~s$^{-1}$) & log($\Phi_\mathrm{H\alpha}^\ast$/Mpc$^{-3}$) & $\Phi_\mathrm{H\alpha=E42.2}$/$\Phi_\mathrm{S13}$ & log(M$_\star^\ast$/M$_\odot$) & log($\Phi_{\mathrm{M}_\star}^\ast$/Mpc$^{-3}$) & $\Phi_\mathrm{M_\star=E9.4}$/$\Phi_\mathrm{D15}$ \\
\hline
All        & $43.466\pm0.120$ & $-2.056\pm0.102$ & 14.1 & $10.955\pm0.178$ & $-2.464\pm0.112$ & 14.0 \\
Group      & $43.937\pm0.441$ & $-2.079\pm0.309$ & 26.7 & $11.505\pm0.588$ & $-2.380\pm0.341$ & 31.6 \\
Intergroup & $43.207\pm0.110$ & $-2.105\pm0.109$ &  8.5 & $10.603\pm0.179$ & $-2.529\pm0.145$ &  7.5 \\
\hline
\end{tabular}
\label{tab3}
\end{table*}

Figure~\ref{fig9} shows the distribution functions for \ha\ luminosity and stellar 
mass of HAEs. As described above, we employed the combined sample of HAEs and 
HAE candidates, and we also show only the HAE in each panel. We have performed the 
non-linear least chi-square fitting by the Schechter function to the entire HAEs, 
and HAEs in group and intergroup environments, with the {\sc mpfit} package 
\citep{Markwardt:2009}\footnotemark[4]. The {\sc mpfit} fitting 
code semi-automatically provides us with the best fitting parameters and errors. We 
remove the faintest and two brightest bins in the fitting process for \ha\ 
luminosity function since the former has a significant completeness issue and the 
the bright-ends may have a large contribution by AGN \citep{Matthee:2017} (and 
indeed the brightest bin includes only RG). Also, we reject the stellar mass bins 
of $<10^{9.4}$ \msun\ for the fitting of stellar mass function since the SFR limit 
causes a critical incompleteness problem for the lower-mass HAEs than this 
threshold along the star-forming main sequence (\S\ref{s3.2}). The sample bins used 
in the curve fitting are highlighted by the thick frames in each panel. 

\footnotetext[4]{\url{http://purl.com/net/mpfit}}

The results of the Schechter function fitting are available in Table~\ref{tab3}. 
The Schechter function is given by the following equations,
\begin{eqnarray}
\phi(L)dL &=& \phi^\ast \bigg(\frac{L}{L^\ast}\bigg)^\alpha \exp\bigg(-\frac{L}{L^\ast}\bigg)\frac{dL}{L^\ast} ~~~~~~~~~~~~~~~~ or \label{eq8} \\
\phi(L)dL &=& \phi^\ast \bigg(\frac{L}{L^\ast}\bigg)^{\alpha+1} \exp\bigg(-\frac{L}{L^\ast}\bigg)\ln{10}~d(\log{L}), \label{eq9}
\end{eqnarray}
where $L$ is \ha\ luminosity or stellar mass in this work. $L^\ast$ is the 
characteristic luminosity or stellar mass in which the power-law slope cuts off. 
We count the number of galaxies as follows, 
\begin{equation}
\phi(\log{L})=\Sigma_i(V_\mathrm{max}\cdot C(NB)\cdot\Delta(\log{L}))^{-1},
\label{eq10}
\end{equation}
where $V_\mathrm{max}$ (co-Mpc$^3$) is the volume size derived from the filter FWHM 
of the narrowband filter (45 co-Mpc) and survey area (58 co-Mpc$^2$). Those of the 
group and the intergroup regions can be estimated to be 609 and 3802 co-Mpc$^3$, 
respectively. Actual volume size should be smaller than those values since HAEs are 
concentrated around the protocluster centre according to \citet{Shimakawa:2014}. We 
ignore this systematic error since we solely compare the shape of distribution 
functions between higher and lower densities within the same protocluster region. 
Due to the small sample size, we have to fix the power-law slope $\alpha=-1.6$ 
for the \ha\ luminosity function and $\alpha=-1.5$ for the stellar mass 
function, respectively, to minimise the fitting errors. These $\alpha$ values 
are employed from \citet{Sobral:2013} and \citet{Davidzon:2017}, respectively. 
In either case, we also tried differing $\alpha$ by $\pm0.1$, and the result did 
not improve. We should keep in mind that our restricted sample sizes may be 
insufficient to determine the robust function parameters and even errors. For 
example, even if one has sufficient statistics on large datasets, there are 
variations of obtained $M_\star^\ast$ and $\Phi^\ast$ due to sample variance and 
selection effects 
\citep{Ilbert:2013,Muzzin:2013,Sobral:2014,Davidzon:2017,Hayashi:2017b}. 
We believe that systematic comparisons between different environments given only 
our data would work well. To compare our samples with results from the literature 
would require far more attention as they have used different sample selections 
based on the different datasets. 

To make fairer comparisons in various environments, we also show both \ha\ 
luminosity and stellar mass functions normalised by those faint-end values of 
the employed sample bins, $\log$($L_\mathrm{H\alpha}$/erg~s$^{-1}$)=42.2 and 
$\log$(M$_\star$/M$_\odot$)=9.4, respectively. These scaled distributions appear in 
the bottom of Fig.~\ref{fig9}, and scaling values can be found in Table~\ref{tab3} 
that would be useful to quantify the overdensities of our targets more generally. 

The derived distribution functions of the two environments show significant 
differences in the sense that the dense groups tend to host a larger number of more 
massive and more active HAEs than the intergroup regions within the same 
protocluster field. The results suggest that environmental effects on the scale of 
$<1$ ph-Mpc change the physical process or time-scale of galaxy formation in this 
young protocluster. Both $L_\mathrm{H\alpha}^\ast$ and M$_\star^\ast$ in HAEs in the 
dense group cores are significantly higher than those in the intergroup regions by 
0.7 and 0.9 dex, respectively. When we compare the \ha\ luminosity functions with 
that for the random field \citep{Sobral:2013}, the dense groups exhibit a much 
larger number of HAEs at higher luminosities, and there may be a slight excess in 
the intergroup environments. According to the comparison of the scaled fitting 
curves, for example, the dense groups host by eight times larger number of HAEs at 
M$_\star=10^{11}$ \msun\ than the intergroup regions. 

On the other hand, interestingly, the intergroup regions likely 
show the very similar shape of stellar mass function as that of star-forming 
galaxies in the general field at similar redshifts \citep{Davidzon:2017}. Not only 
derived mass function, but also those stellar mass bins are well consistent with 
\citet{Davidzon:2017}, suggesting non-environmental dependence of stellar mass 
distribution outside of the densest regions despite that the intergroup regions are 
also associated with the same dense protocluster, and have eight times higher 
densities on average than the general fields. However, we should note that stellar 
mass function of \citet{Davidzon:2017} has a lower cutoff than those of the past 
studies \citep{Ilbert:2013,Muzzin:2013,Sobral:2014}, and thus our comparison result 
changes depending on the literature and the power slope ($\alpha$) used for the 
line fitting. To resolve this issue, more comprehensive analyses in both 
protoclusters and general fields should be required. 

On top of that, the faint end profiles of distribution functions should be another 
intriguing point. Environmental dependence of stellar mass function in the low-mass 
end at low redshifts has been widely studied 
\citep{Trentham:1998,Conselice:2002,Yagi:2002,Propris:2003,Penny:2008}, 
especially toward a better understanding of the environmental quenching 
\citep{Peng:2010,Bolzonella:2010,Mortlock:2015}.
At higher redshift, $z\sim1$, \citet{Sobral:2011} and 
\citet{Tomczak:2017} find the flatter slope at the faint end in higher local 
overdensities. 
However, the faint end slope of star-forming galaxies in high-$z$ protoclusters at 
$z>2$ has been poorly understood. \citet{Cooke:2014} have reported a steeper $\alpha$ 
in the stellar mass distribution in protocluster regions based on the Millennium 
Simulation \citep{Springel:2005}. On the other hand, as also mentioned by 
\citet{Cooke:2014}, high densities could involve environmental quenching in the large 
host haloes at even such high redshifts. 
Regrettably, the current datasets cannot constrain the power-law slope $\alpha$ for 
both \ha\ and stellar mass functions due to the small sample size. However, judging 
from the comparisons of the normalised distribution functions, we cannot see the 
clear distinctions from each other at the faint end of these functions. The much 
larger datasets must be needed to improve the statistics and quantify the faint end 
slopes more precisely.


\subsection{Environmental dependence within the protocluster}\label{s3.5}

We here summarise the comparison tests of physical properties between HAEs in lower 
and higher densities selected based on the 5th mean projected distance 
($\overline{b}_\mathrm{5th}\gtrless220$ ph-kpc) across the survey field. We test 
if there are any statistical differences between two groups by using the KS test 
in nine properties: (a) stellar mass, (b) SFR, (c) sSFR, (d) deviation from the 
main sequence ($\Delta\mathrm{MS}$), (e) dust reddening 
derived from the SED-fitting, (f) offset from the stellar mass--E(B$-$V) relation 
($\Delta$E(B$-$V), see Fig.~\ref{fig4}), (g) effective radius, (h) deviation from 
the mass--size relationship by \citet{Wel:2014} ($\Delta\mathrm{Re_{V14}}$), and 
(i) J $-$ Ks colour striding over Balmer/4000\AA\ break.

\begin{figure*}
\centering
\includegraphics[width=0.9\textwidth]{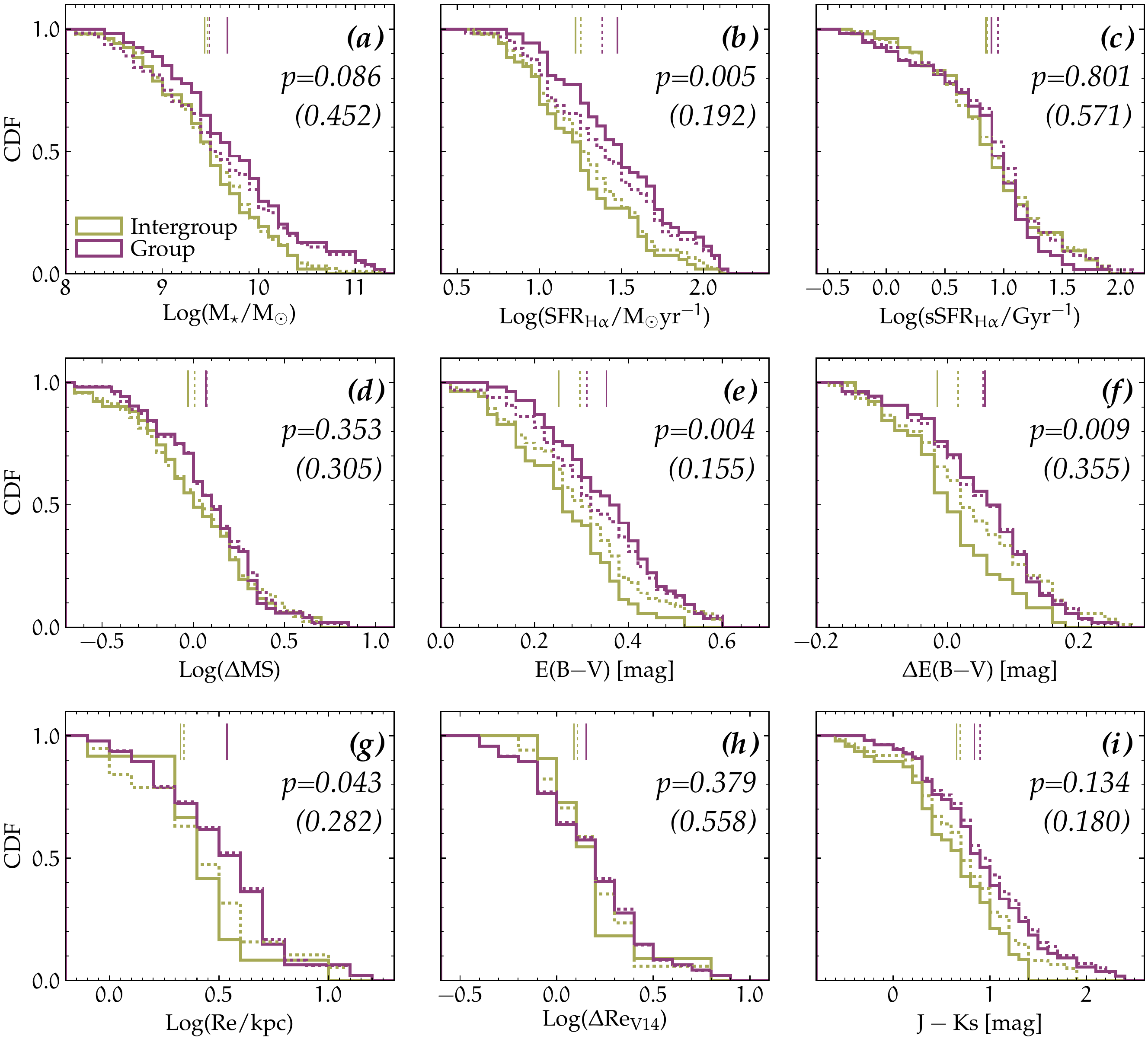}
\caption{Cumulative distribution functions (CDFs) for galaxies properties, (a) 
stellar mass, (b) SFR, (c) sSFR, (d) deviation from the main sequence, 
$\Delta\mathrm{MS}$, (e) SED-inferred stellar extinction 
E(B$-$V), (f) offsets from the mass--E(B$-$V) sequence of the entire HAEs, (g) 
effective radius, $R_e$, (h) offset from the mass-size relation of star-forming 
galaxies by \citet{Wel:2014} ($\Delta\mathrm{Re_{V14}}$), and (i) J$-$Ks colour 
striding Balmer/4000\AA\ break. Purple and yellow lines indicate CDFs of HAE in 
group region and intergroup region, which are divided by the median values of the 
mean projected separation of 220 ph-kpc. Solid lines show the HAE sample and 
dotted lines include HAE candidates. The vertical lines in each panel show the 
median values of respective samples (Table~\ref{tab4}).}
\label{fig10}
\end{figure*}

\begin{figure*}
\centering
\includegraphics[width=0.9\textwidth]{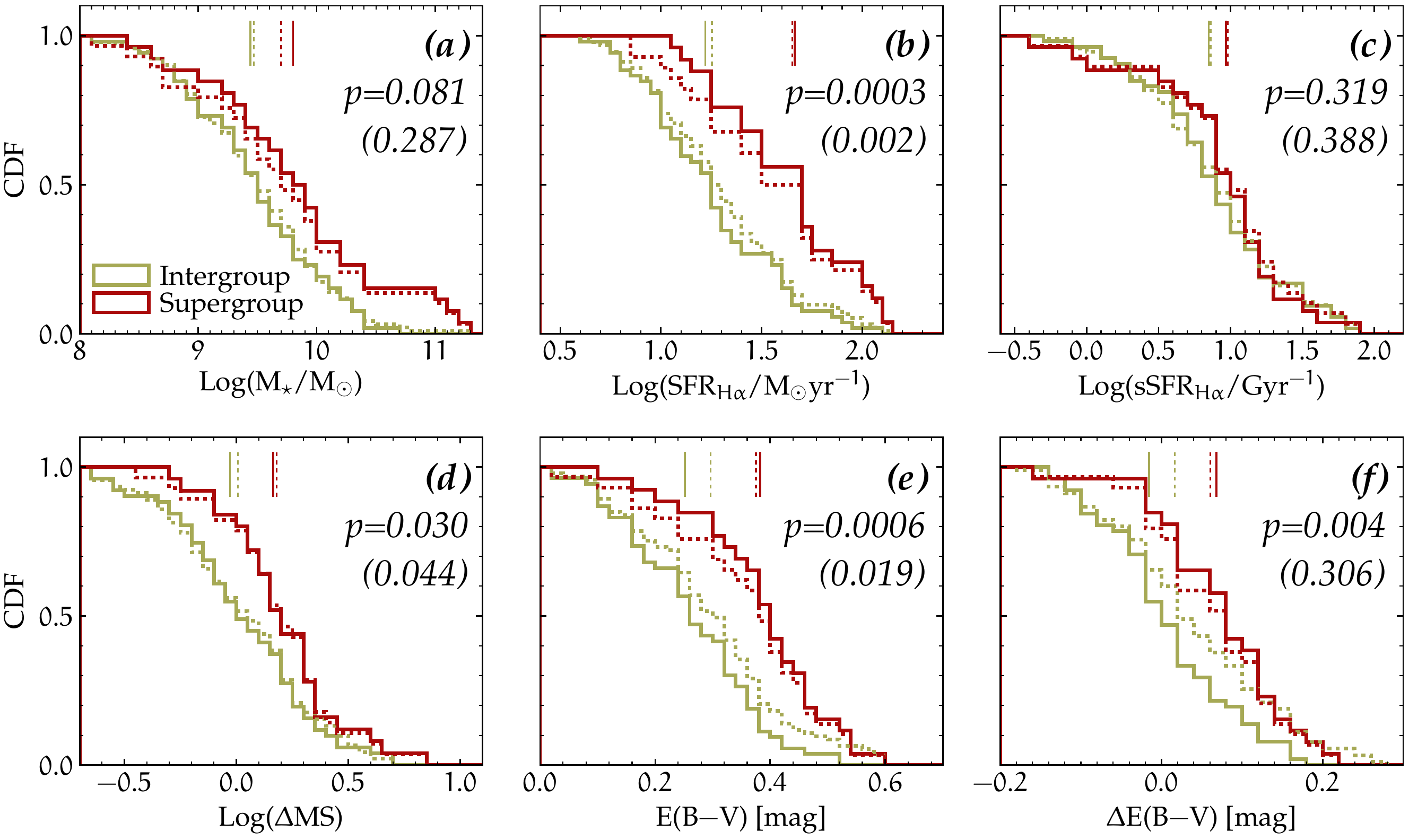}
\caption{Cumulative distribution functions (CDFs) for galaxies properties, (a) 
stellar mass, (b) SFR, (c) sSFR, (d) $\Delta\mathrm{MS}$, (e) E(B$-$V), and 
(f) $\Delta$E(B$-$V) of HAEs in the supergroup (reds) and the intergroup (yellow) 
regions. The solid line shows the HAE sample and dotted line include HAE candidates. 
The vertical lines in each panel show the median values of respective samples.}
\label{fig11}
\end{figure*}

\begin{table*}
\caption{The results of the KS test for nine physical parameters listed in the first 
column. The second--fourth columns show median values of each parameter of HAEs in 
the group (50th \%tile higher densities), the intergroup (50th \%tile lower densities), 
and the supergroup regions (top quartile higher densities), respectively. 
The fifth and sixth columns indicate $p$-values of the KS test for each physical 
parameter. Values enclosed in parentheses are based on the sample including HAE 
candidates.}
\begin{center}
\begin{tabular}{lccccc}
\hline
Parameter & $\mathrm{\tilde{Group}}$ & $\mathrm{\tilde{Intergroup}}$ & $\mathrm{\tilde{Supergroup}}$ & $p_\mathrm{(Group-Intergroup)}$ & $p_\mathrm{(Supergroup-Intergroup)}$ \\
\hline
log(M$_\star$/M$_\odot$)       & 9.68 (9.49) &    9.44 (9.47) & 9.80 (9.70) & 0.086 (0.452) & 0.081 (0.287) \\
log(SFR/M$_\star$~yr$^{-1}$)   & 1.48 (1.38) &    1.22 (1.25) & 1.66 (1.65) & 0.005 (0.192) & 0.0003 (0.002) \\
log(sSFR/Gyr$^{-1}$)           & 0.90 (0.95) &    0.85 (0.86) & 0.97 (0.98) & 0.801 (0.571) & 0.319 (0.388) \\
log($\Delta$MS)                & 0.07 (0.07) & $-$0.03 (0.01) & 0.16 (0.18) & 0.353 (0.305) & 0.030 (0.044) \\
E(B$-$V)                       & 0.35 (0.31) &    0.25 (0.30) & 0.38 (0.38) & 0.004 (0.155) & 0.0006 (0.019) \\
$\Delta$E(B$-$V)               & 0.06 (0.05) & $-$0.02 (0.02) & 0.07 (0.06) & 0.009 (0.355) & 0.004 (0.306) \\
log(Re/kpc)                    & 0.54 (0.54) &    0.33 (0.34) & 0.55 (0.55) & 0.043 (0.282) & 0.049 (0.334) \\
log($\Delta$Re$_\mathrm{V14}$) & 0.15 (0.15) &    0.09 (0.11) & 0.14 (0.14) & 0.379 (0.558) & 0.468 (0.769) \\
J$-$Ks                         & 0.84 (0.90) &    0.66 (0.82) & 0.92 (0.99) & 0.134 (0.180) & 0.100 (0.102) \\
\hline
\end{tabular}
\end{center}
\label{tab4}
\end{table*} 

Obtained $p$-values from the KS test and median values in each physical parameter 
are available in Table~\ref{tab4} and Fig.~\ref{fig10}. Table~\ref{tab4} also shows 
the results of the KS test between HAEs in the supergroup and the intergroup 
regions. The comparisons of those cumulative distribution functions can be seen in 
Fig.~\ref{fig11} for the former six parameters that show different $p$-values as 
compared to those between the group and the intergroup regions. These results 
address environmental effects for galaxy formation in the extreme. The comparison 
results show statistically different between group and intergroup HAEs in SFR, 
E(B$-$V), and effective radius, with $p$-values lower than 5 per cent. 
These suggest that HAEs in the dense groups tend to have higher SFRs, larger 
amounts of dust extinctions, and larger stellar sizes, respectively. As mentioned 
in \S\ref{s3.2} and \S\ref{s3.3}. 

However, such trends seen in SFR and effective radius can no longer be confirmed 
for a given stellar mass, according to the comparisons for $\Delta\mathrm{MS}$ and 
$\Delta\mathrm{Re_{V14}}$, respectively. Indeed, such SFR and size differences 
between group and intergroup regions are supposed to be caused by the significant 
excess of massive HAEs in the dense cores, as provided by these stellar mass 
function (Fig.~\ref{fig9}). 
On the other hand, the statistical difference seen in the dust reddening of HAEs 
remains consistent with $p=0.009$ if we compare those for a given stellar mass 
($\Delta$E(B$-$V) in Table~\ref{tab4}) along the fitted line of the 
mass--extinction relation (Fig.~\ref{fig4}). Such higher dust obscuration of 
galaxies in local overdensities for a given stellar mass range is consistent with 
the results of past studies \citep{Koyama:2013b,Hatch:2017} that have investigated 
the environmental dependence of dust reddening of galaxies on a more global scale. 
However, our measurement of the dust reddening is based on the SED-fitting under 
the assumption of fixed stellar metallicity and limited star formation history. 
Therefore, such a systematic difference could be also interpreted by difference in 
metal abundance, age, and tau between the two samples, which is actually suggested 
by the previous studies (e.g. \citealt{Mateu:2014,Genel:2016}). 
It should be noted that at least we could not find clear differences in the 
SED-inferred age and tau between HAEs in higher and lower densities. To resolve 
this problem, we need more direct estimation, e.g. Balmer decrement technique based 
on \ha\ and \hb\ line spectra \citep{Shivaei:2015}. 

Furthermore, HAEs in extremely high (top quartile) overdensities categorised as 
the supergroup, show more discriminable properties as compared to those in the 
intergroup regions. The KS test shows even lower $p$-values in the comparisons of 
most physical properties, and we for the first time identify that the supergroup 
HAEs tend to have higher SFRs across the star-forming main sequence than the 
intergroup HAEs. As inferred from the margin of those median $\Delta\mathrm{MS}$ 
values of 0.19 dex, stellar mass assembly in the supergroups are by 1.5 times 
accelerated as a function of stellar mass, which may largely contribute to the 
excess of massive HAEs in local overdensities seen in the stellar mass functions. 
Such a prominent deviation of the main sequence in higher densities has not been 
found in the past studies \citep{Koyama:2013a,Koyama:2013b}. They have 
investigated the environmental dependence of star-forming activities on a more 
global scale based on shallower data than this paper. Such different observing 
scale and survey depth may be one of the primary reasons of the discrepancy. 
Indeed, our previous shallower data in the same USS~1558 field show no 
significant $\Delta\mathrm{MS}$ enhancement in the densest group region 
\citep{Hayashi:2012}, suggesting sufficiently deep \ha\ imaging is important to 
identify enhanced star formation in the extremely dense groups. 
Moreover, differences in the forming phase of the clusters can be another key 
factor, since for example, our target USS~1558 at $z=2.5$ is apparently younger 
protocluster system than PKS~1138$-$262 at $z=2.2$ studied by 
\citet{Koyama:2013a,Koyama:2013b} in the sense of less established red sequence, 
more clumpy substructure, and higher redshift in USS~1558 relative to PKS~1138. 
We discuss more detailed explanations on the physical origins of the SFR 
enhancements in the discussion section (\S\ref{s4}). 

Lastly, we cannot identify the clear difference in the other physical 
properties shown in Fig.~\ref{fig10}. First, this work has found the significant 
enhancement of massive HAEs in the dense cores. However, mass distributions are 
consistent within errors between the group and the intergroup regions at the 
stellar mass lower than $10^{10.5}$ \msun, and thus, the difference of mass 
distributions is no longer significant in the entire mass range. Also, we cannot 
rule out the fact that size distributions do not depend on local overdensities 
for a given stellar mass because of high $p$-values ($>0.05$) in the comparisons 
of $\Delta\mathrm{Re_{V14}}$ (see also \S\ref{s3.3}). This means that early 
environmental effects in the protocluster may solely affect the stellar mass 
assembly histories (and probably dust reddening) of galaxies, but have little 
impact on those size growth histories. Besides, our results show no clear 
difference in Balmer/4000\AA\ colour (J$-$Ks). In USS~1558, a tight red sequence 
is not established yet unlike the more advanced protocluster systems at $z\sim2$ 
\citep{Kodama:2007}, which may be the reason why we cannot see the significant 
excess of redder HAEs in the dense cores.


\section{Discussion}\label{s4}

Using a total of 107 HAE sample and 51 additional HAE candidates, this work has 
analysed the environmental dependence of physical properties of \ha-emitting 
galaxies within the protocluster on the substructure ($<1$ ph-Mpc) scale. Despite 
such a small scale, we find a significant enhancement of more massive and active 
HAEs in the densest parts of the protocluster. We discuss potential physical 
mechanisms that drive these observational tendencies and try to explain galaxy 
formation in young star-forming protoclusters like our target USS~1558, together 
with previous observational and theoretical work.

\begin{figure}
\centering
\includegraphics[width=.9\columnwidth]{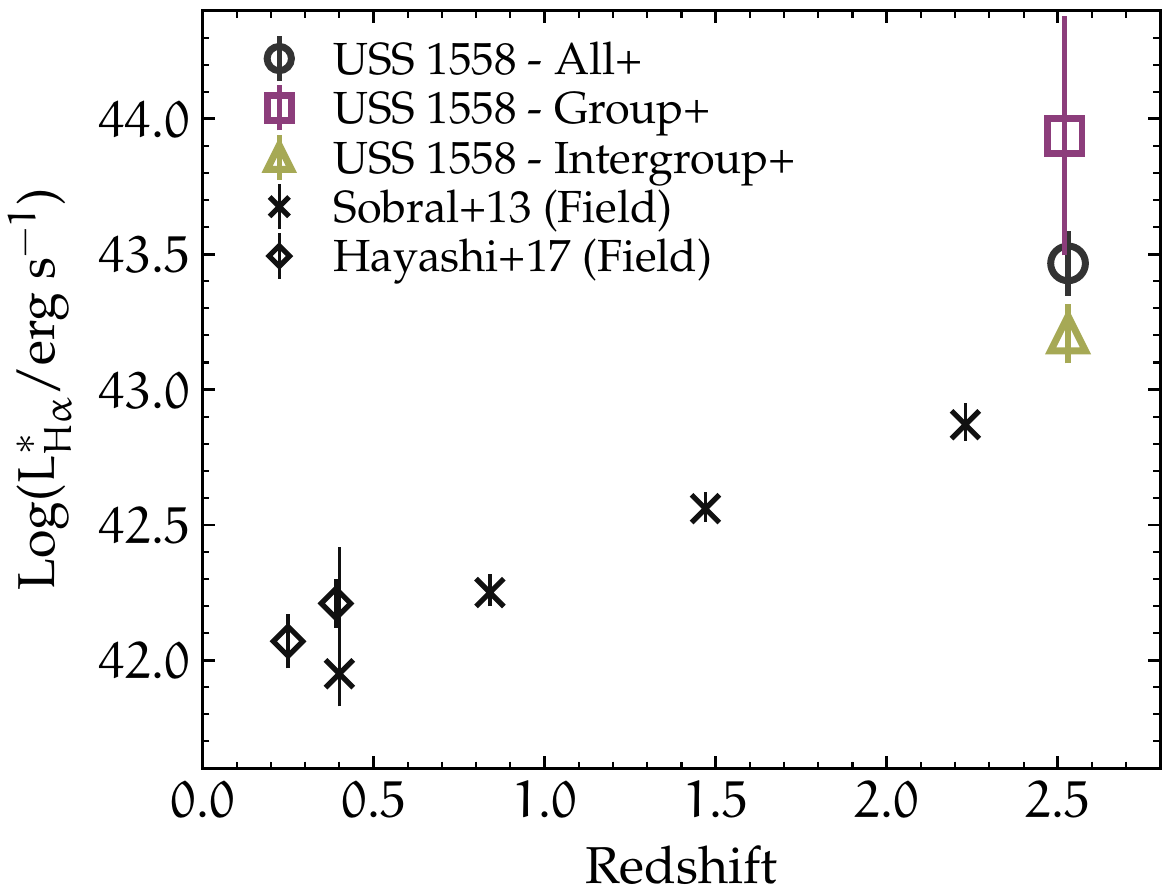}
\caption{The characteristic \ha\ luminosity ($L_\mathrm{H\alpha}^\ast$) as a function 
of redshift. Purple square, black circle, and yellow triangle are based on our sample 
in the entire protocluster, group, and intergroup regions in USS~1558 at $z=2.53$, 
respectively. The purple square symbol is a bit shifted leftward to improve 
visualisation. Crosses and diamonds are $L_\mathrm{H\alpha}^\ast$ derived from the 
similar \ha\ survey for the general fields at lower redshifts by \citet{Sobral:2013} 
and \citet{Hayashi:2017b}, respectively. Errorbars show $1\sigma$ fitting errors.}
\label{fig12}
\end{figure}

Firstly, the comparison of cumulative distribution functions reveals a statistical 
difference in SFRs of HAEs in between lower and higher densities, which would be 
independent of uncertainties of estimation of dust reddening since this tendency 
still can be confirmed without dust correction. Figure~\ref{fig12} represents 
the characteristic \ha\ luminosities ($L_\mathrm{H\alpha}^\ast$) as a function of 
redshift for our samples and field galaxies \citep{Sobral:2013,Hayashi:2017b}. 
These are derived based on the Schechter function assuming a fixed $\alpha=-1.6$, 
which allows us to make a relatively fair comparison of the cutoff \ha\ luminosities 
of the power-law form among these samples. Also, one should note that all of the 
reference samples are based on narrowband imaging surveys. 

As seen in the figure, HAEs in the intergroup region of USS~1558 follow the 
extrapolation of the redshift evolution of the characteristic \ha\ luminosity 
reported by \citet{Sobral:2013}, while HAEs in dense group environments deviate from 
the sequence. This suggests that the densest environment in the young protocluster 
have a larger number of active star-forming galaxies than the intergroup regions 
in the same protocluster field. Such higher cutoff \ha\ luminosity in overdensities 
were not explored in \citet{Hatch:2011} for a protocluster PKS~1138$-$262 at $z=2.2$. 
However, the additional \ha\ imaging for the same field has found a number excess of 
massive HAEs with higher SFRs along the star-forming main sequence 
\citep{Koyama:2013a,Koyama:2013b}, suggesting that this trend might be a common 
feature during rapid building of bright red sequence objects in the protocluster. 
We should note, however, environmental dependence of galaxy properties will depend 
on their evolutionary phase of each protocluster system. 

The bright-end HAEs in \ha\ luminosity function may host active 
galactic nuclei (AGN). \citet{Sobral:2016b} find a tight correlation between \ha\ 
brightness and AGN fraction, which for example shows 50 per cent AGN fraction in 
HAEs with $L_\mathrm{H\alpha}=5\times L_\mathrm{H\alpha}^\ast$ (see also 
\citealt{Matthee:2017}). Thus, taking account of the potential increase of AGN 
fraction in high-$z$ protoclusters \citep{Lehmer:2009,Lehmer:2013}, we could explain 
the higher cutoff value of \ha\ lunminosity function in the dense groups by a 
substantial excess of AGN. 
However, we see the number excess of HAEs near $L_\mathrm{H\alpha}^\ast$ and such a 
significantly luminous HAE ($L_\mathrm{H\alpha}\sim10^{44}$ erg~s$^{-1}$) 
is only the radio galaxy in our sample. Thus, it may seem to be difficult to examine 
the higher characteristic \ha\ luminosity only by the substantial excess of AGNs in 
the group environments. Future deep follow-up X-ray and near-infrared spectroscopy 
will help us to test the connection between AGN activities and environmental 
dependence of the luminosity function. 

We also identify a prominent enhancement of massive HAEs (M$_\star>10^{10.5}$ 
\msun) in the dense groups by comparing those stellar mass distribution with that of 
the intergroup HAEs. This is broadly consistent with 
\citet{Steidel:2005,Koyama:2013a,Koyama:2013b} who reported similar trends by 
comparing protocluster and field galaxies, while we find the excess of massive HAEs 
in substructures within the protocluster. It is crucially important to understand 
characteristics of those massive \ha-emitting galaxies to infer the physical origins 
of the strong environmental dependencies seen in the local Universe, 
since they would be the progenitors of early-type galaxies seen at the bright end of 
the red sequence seen in today's galaxy clusters 
\citep{Gladders:2000,Thomas:2005,Koester:2007,Stott:2009,Cappellari:2011}. 
However, the current relevant datasets seem to argue that most of massive HAEs except 
the RG are likely star-forming galaxies rather than AGNs, for example, because those 
rest optical line spectra available in five out of six objects do not show strong 
\nii\ emission line features \citep{Shimakawa:2015b}. 
With this motivation, we are now working on various follow-up observations in the 
multi-wavelength regime. Upcoming results from these surveys will help us to 
characterise these massive HAEs in the near future. 

On the other hand, HAEs outside of the dense cores have similar SFR and stellar mass 
distributions to those in the random field \citep{Sobral:2013,Davidzon:2017}. These 
suggests that environmental impacts are not critical in intermediate densities 
in young protoclusters i.e. an early stage of inside-out growth of the galaxy 
clusters. This also means that high sampling density enough to resolve substructures 
on smaller than $\sim1$ Mpc scale should be needed to identify early environmental 
effects as represented by our deep narrowband survey. If not, poor sampling density 
would lead us to misinterpret obtained results despite the existence of some 
important features hidden in the densest substructures in protoclusters. 

The total SFR densities estimated from integration of the \ha\ luminosity function 
down to $L_\mathrm{H\alpha}=1\times10^{42}$ erg~s$^{-1}$ is 12 
\msun~yr$^{-1}$Mpc$^{-3}$ and 1111 \msun~yr$^{-1}$Mpc$^{-3}$ in the intergroup and 
the group regions, corresponding to two and four orders of magnitude higher SFR 
volume densities than that in the general field 
\citep{Madau:1996,Lilly:1996,Madau:2014} at the same redshift, respectively. Such 
high cosmic SFR densities in dense cores of protoclusters will rapidly decline from 
$z\sim2$ to the local Universe by the scale of $\propto(1+z)^{6-7}$ 
\citep{Kodama:2004,Finn:2005,Cooper:2008,Koyama:2010,Koyama:2011,Hayashi:2011,Webb:2013,Smail:2014,Clements:2014,Shimakawa:2014,Kato:2016}. 
This drastic decrease of star-formation densities is intimately related to a rapidly 
increasing fraction of quenching populations 
\citep{Baldry:2006,Peng:2010,Sobral:2011,Muzzin:2012,Darvish:2016} 
driven perhaps by cold gas depletion (strangulation), energy injections of member 
galaxies and AGNs, and environmental quenching (e.g. 
\citealt{Jaffe:2015,Noble:2016,Maier:2016,Valentino:2016,Wang:2016,Hayashi:2017}, 
and simulations by e.g. \citealt{Romeo:2005,Sijacki:2007,Fabjan:2010,Vogelsberger:2014}). 
In the meanwhile, star-forming activity propagates towards the outskirts and in 
smaller systems in the context of inside-out growth of the galaxy 
clusters \citep{Poggianti:2006,Elbaz:2007,Koyama:2011,Burg:2015,Chiang:2017}. 
Eventually, the picture is entirely reversed in the sense that galaxies in massive 
clusters and groups tend to have lower SFRs than the general field in the 
present-day galaxy clusters and dense group environments as successfully 
represented by the Sloan Digital Sky Survey 
(\citealt{Lewis:2002,Gomez:2003,Kauffmann:2004}, see also \citealt{Alberts:2014}). 

Moreover, the other intriguing feature discovered by this paper is the 
identification of enhanced SFRs {\it across} the main sequence in very high local 
densities. So far, while the past similar studies have found enhanced SFRs 
{\it along} the star-forming main sequence due to the number excess of massive HAEs 
\citep{Koyama:2013a,Koyama:2013b}, these massive HAEs are shifted right upward on 
the M$_\star$--SFR plane nearly alongside of the main sequence and do not show 
deviation from the sequence (see also \citealt{Brodwin:2013}). In fact, we do not 
identify any significant difference of $\Delta\mathrm{MS}$ between the group and the 
intergroup HAEs, and only see a deviation when we focus on the top quartile of 
overdensities. This suggests that relatively large sample size enough to trace 
substructures may be important to identify the significant difference in 
such scattered empirical relationship. Combined with mass enhancement in the group 
regions, the densest regions in young protoclusters would have had accelerated and 
enhanced galaxy formation, which has been implied in the past studies for 
(proto)clusters of galaxies at lower redshifts (e.g. 
\citealt{Steidel:2005,Thomas:2005,Strazzullo:2010,Tanaka:2013b,Mateu:2014,Cerulo:2016}, 
but see \citealt{Thomas:2010,Gobat:2013}). 

However, we have no clear explanation of physical origins of the enhanced 
star formation of HAEs in the extreme overdensities. We here propose two scenarios 
which may be able to examine the enhanced star formation of galaxies in the densest 
structures in the young protocluster: (1) galaxy--galaxy interaction and (2) a vast 
amount of cold gas associated with the densest substructures, as described below. 

(1) Because of very high densities no larger than $\overline{b}_\mathrm{5th}=150$ 
ph-kpc in the mean projected separations, it may be expected that there are lots of 
physical interactions among protocluster members therein. In high-density regions, 
galaxy collisions would occur more often as predicted theoretically 
\citep{Okamoto:2000,Gottlober:2001,Fakhouri:2009,Jian:2012}, and also observationally 
(\citealt{Lin:2010,Kampczyk:2013,Hine:2016}, but see \citealt{Darg:2010}). This helps 
protocluster galaxies to grow into the compact, spheroidal galaxies seen in the local 
galaxy clusters through dissipational processes 
\citep{Mihos:1996,Barnes:1996,Bournaud:2007,Jesseit:2009,Hopkins:2008,Hopkins:2010}. 
Indeed, there are several merger-like systems in our HAE sample identifiable by 
visual inspection of F160W/WFC3 images from HST; however, it is hard to classify 
those as mergers or clumpy galaxies based only on rest-optical images. Follow-up 
integral field spectroscopy would constrain their kinematics. Furthermore, 
merger-induced starbursts are likely stochastic events and may not produce a 
systematic upward offset from the star-forming main sequence. Taking account these 
issues, we disfavour the primary explanation by galaxy--galaxy interactions. 

(2) Circumgalactic/intergalactic media (CGM/IGM) are critical components for galaxy 
formation. This is crucially linked to the gas feeding mechanism of galaxies that 
depends on the cosmic age, halo mass and environment \citep{Keres:2005,Voort:2011}. 
At high redshifts ($z\gtrsim2$), as reported by the recent hydrodynamic simulations,  
more efficient radiative cooling and self-shielding at high redshifts allow gas to 
cool and penetrate into galaxies in even hot massive halos larger than 
$1\times10^{12}$ \msun\ since those still hold a significant amount of cold \hi\ gas 
(\citealt{Dekel:2006,Dekel:2009,Dekel:2009b,Keres:2009,Faucher:2010,Voort:2011}, see 
also more classical picture in \citealt{Gunn:1972}). 
The vast amount of \hi\ gas associated with the young protocluster cores might be 
able to examine \lya\ depletion effect and more extended diffuse \lya\ haloes in 
overdensities \citep{Matsuda:2012,Hennawi:2015,Shimakawa:2017b}, and \hi\ column 
density distribution at high redshifts \citep{Voort:2012}. On a large scale, at 
least, recent observations have revealed that massive \hi\ structures are associated 
with protoclusters \citep{Lee:2016,Cai:2017b,Mawatari:2017}. 
Furthermore, cold accretion into the hot cores as known as the cold stream 
\citep{Dekel:2009}, not only help massive galaxies to maintain their star-forming 
activities but also could supply surrounding members sharing the dense group cores 
with sufficient cold gas reservoirs. This might involve more active star formation 
and then produce more dust in the entire HAEs in the densest groups. If true, such a 
vigorous gas feeding mechanism can systematically boost star formation of galaxies 
associated with the young protocluster's core. To test this hypothesis, we desire 
additional observational constraints from another aspect (e.g. CO observation of the 
protocluster centre; \citealt{Emonts:2016}). 

Finally, we acknowledge every result reported in this paper is based on only one 
dense protocluster system at $z=2.53$. A more comprehensive and systematic analyses 
is highly desirable to establish a consensus view on galaxy formation in the galaxy 
cluster. For instance, more than a thousand galaxies or ten dense protoclusters 
would be preferable to significantly improve the derivation of the stellar mass 
function in the protocluster region. On top of that, ideally, we need to investigate 
both star-forming and passive galaxies in galaxy clusters across cosmic time, and 
then build a self-consistent model of quenching histories of cluster galaxies. 
On-going extensive programs such as MAMMOTH (MApping the Most Massive Overdensity 
Through Hydrogen, \citealt{Cai:2016}) and the Subaru Strategic Program with Hyper 
Suprime-Cam \citep{Aihara:2017} have great potential to construct a large 
protocluster sample to address this problem in the near future.



\section{Summary}\label{s5}

\begin{description}

\item[---]
Deep \ha\ line imaging of the protocluster, USS~1558 at $z=2.53$ with MOIRCS on the 
Subaru Telescope has succeeded in discovering a total of 107 HAEs within just a 
$4'\times7'$ square arcmin area. Stellar masses and dust-corrected SFRs are measured 
down to $1\times10^8$ \msun\ and 3 \msun~yr$^{-1}$, respectively. 
The high sampling density by deep panoramic \ha\ mapping enables the analyses of 
early environmental effects of galaxy properties within the protocluster, on 
substructure scales smaller than one ph-Mpc. We derived a density peak with galaxies 
by the mean projected separation less than 100 ph-kpc scales. We then divide the 
HAEs into sub-samples by overdensity to compare physical properties with local 
environment in the protocluster. 

\item[---]
This work compared physical properties of HAEs between lower and higher 50th 
percentile densities within the protocluster. We find enhancements of SFRs and dust 
reddening of HAEs in the dense group cores. At a given stellar mass, we only see 
statistical differences in the dust reddening between the group and the intergroup 
regions. However, we find that HAEs in the supergroup regions (top quartile 
overdensity) show significant SFR enhancements across the main sequence. Moreover, 
we have identified that \ha\ luminosity and stellar mass distributions are 
remarkably different between the group and intergroup environments in the sense that 
the dense group regions in the protocluster have significantly higher cutoff 
\ha\ luminosity and stellar mass than the intergroup regions. These results suggest 
that the densest environment in the early phase of inside-out growth of the galaxy 
clusters contribute greatly to the rapid building of very massive galaxies. 
Importantly, these unique features can be found when one has sufficiently deep data 
and high sampling density to detect faint objects and resolve substructures in the 
protocluster. 

\item[---]
We argue that the densest parts of the young protocluster take the lead in prompt 
and early construction of very massive galaxies. These will eventually grow into 
the bright-end red sequence of objects commonly seen in present-day massive clusters 
of galaxies. 
Meanwhile, in the early forming protocluster phase, intergroup environments have 
little involvement in building such massive galaxies in accordance with the primary 
stage of inside-out evolution of galaxy clusters. Exploring the kinematics of 
merging protocluster members and focusing on gas feeding processes into 
protocluster's cores is crucial to reveal the physical phenomena that boost those 
star formation.

\end{description}


\section*{Acknowledgements}

The data are collected at the Subaru Telescope, which is operated by the National 
Astronomical Observatory of Japan. A part of analyses is conducted with the assistance 
of the Tool for OPerations on Catalogues And Tables ({\sc topcat}; 
\citealt{Taylor:2015}). This work gains the benefit from the 3D-HST Treasury Program 
(GO 12177 and 12328) with NASA/ESA HST, which is operated by the Association of 
Universities for Research in Astronomy, Inc., under NASA contract NAS5-26555. 
We thank the anonymous referee for useful comments. 
R.S. acknowledge the support from the Japan Society for the Promotion of Science (JSPS) 
through JSPS overseas research fellowships. T.K. acknowledges KAKENHI No. 21340045.




\bibliographystyle{mnras}
\bibliography{bibtex_library} 


\appendix
\section{Catalogue}\label{a1}

\begin{table*}
\begin{center}
\caption{Data catalogue of the HAE sample in USS~1558. Full table is available as 
online material.}
\begin{tabular}{llcccccc}
\hline
ID & $z_\mathrm{spec}$ & $\overline{b}_\mathrm{5th}$ & $\log$(M$_\star$) & E(B$-$V) & $\log$(SFR$_\mathrm{H\alpha}$) & $\log$(R$_e$) & J$-$Ks \\
  &  & (ph-kpc) & (M$_\odot$) & (mag) & (M$_\odot$yr$^{-1}$) & (kpc) & (mag) \\
\hline
0002 &  ---     & 292 & $8.76_{-0.13}^{+0.53}$ & $0.16_{-0.10}^{+0.03}$ & $0.72_{-0.19}^{+0.15}$ 
&      ---      & $0.22\pm0.54$ \\
0003 &  ---$^L$ & 415 & $8.08_{-0.28}^{+1.38}$ & $0.08_{-0.08}^{+0.07}$ & $0.77_{-0.16}^{+0.15}$ 
&      ---      &      ---      \\
0005 &  ---     & 737 & $9.73_{-0.03}^{+0.04}$ & $0.23_{-0.01}^{+0.01}$ & $1.25_{-0.10}^{+0.10}$ 
&      ---      & $-0.37\pm0.12$ \\
     ...     &  ...     & ... &           ...          &           ...          &           ...          
&      ...      &      ...      \\
\hline \\
\multicolumn{8}{l}{$^L$ Dual \ha\ and \lya\ emitters identified by \citet{Shimakawa:2017b}.}
\end{tabular}
\end{center}
\label{tab5}
\end{table*}


\section{Completeness}\label{a2}

\begin{figure}
\centering
\includegraphics[width=0.9\columnwidth]{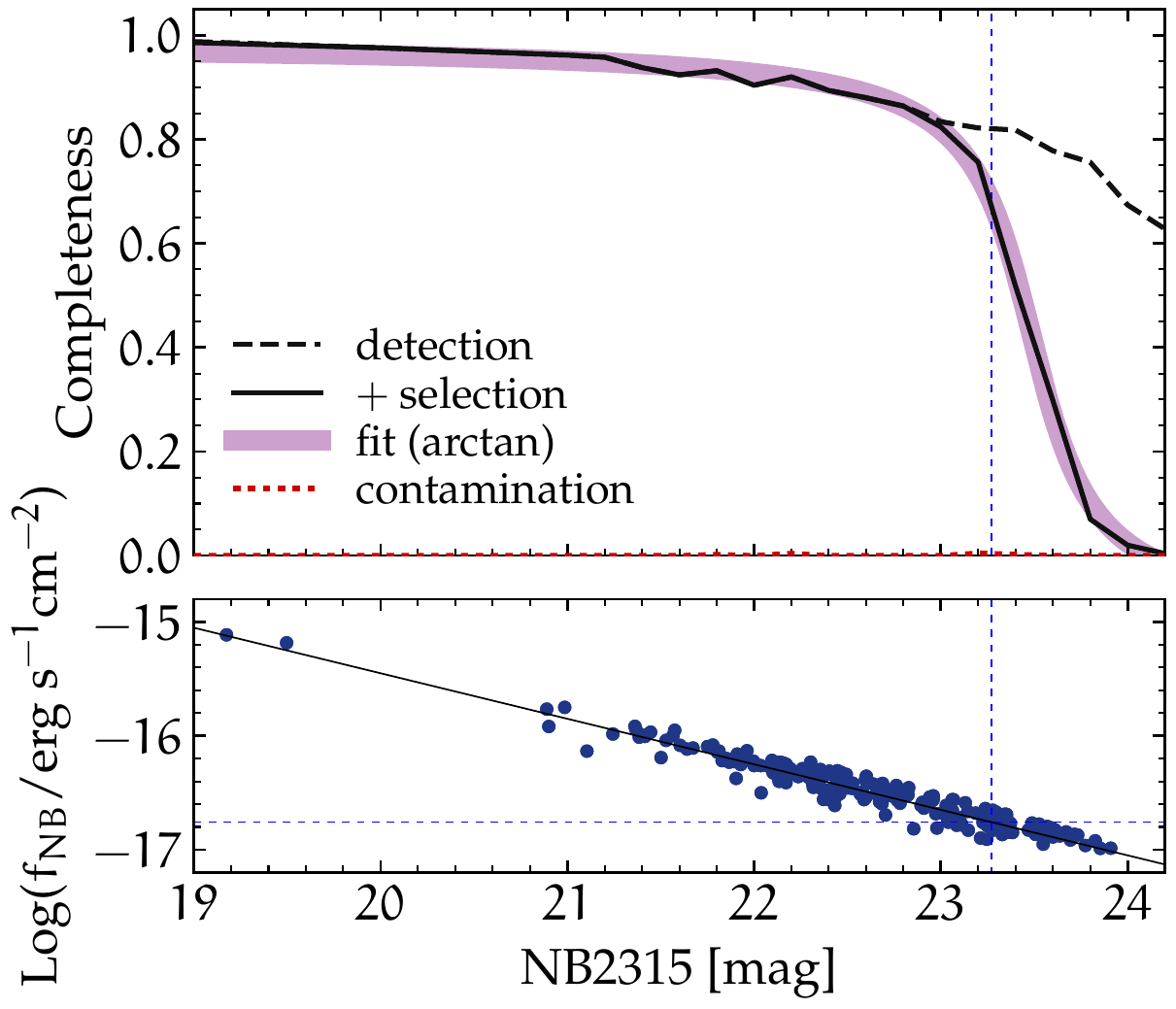}
\caption{Dashed line presents the detection completeness as a function of 
narrowband magnitude, and the solid line includes the selection completeness. We 
use an arctangent function as fitted to the black line by the purple thick curve. 
The red dotted line indicates the contamination rate from photometric errors. 
In the lower figure, blue circles show log narrowband fluxes ($f_{NB}$) of our 
narrowband emitters. The black solid line indicates the fit to the observed emitter 
flux for the Monte Carlo simulation on selection completeness. The blue vertical 
and perpendicular lines indicate 68 per cent completeness, as estimated from our 
simulation.}
\label{fig13}
\end{figure}

We calculate completeness as a function of narrowband magnitude by using a Monte 
Carlo simulation. First, we embedded 10 PSF objects at the narrowband magnitude of 
19, 20, and from 21 to 25 mag in step of 0.2 mag to the reduced $NB2315$ image, and 
then test the recovery rate i.e. detection completeness of sources detected by the 
SExtractor (version 2.19.5; \citealt{Bertin:1996}). This test was repeated 50 
times, and thus a total 500 PSF of objects were computed for a given magnitude. The 
derived detection completeness is represented in Fig.~\ref{fig13}. 

We then demonstrate the recovery rate by the narrowband selection expressed by the 
two equations~\ref{eq1} and \ref{eq2}. 
This called selection completeness \citep{Sobral:2009,Sobral:2012,Sobral:2013} 
provides us with number of objects missed in the narrowband selection due to the 
colour-term variation and photometric errors of the $NB2315$ and $K_s$ bands. 
Addressing the selection completeness is crucial to determine the completeness 
correction of the narrowband-selected emitter sample since the narrowband selection 
depends heavily on both photometric errors of narrowband ($NB2315$) and the 
counterpart broadband ($K_s$) as described in the equation~\ref{eq1} and \ref{eq2}. 

\begin{figure}
\centering
\includegraphics[width=0.9\columnwidth]{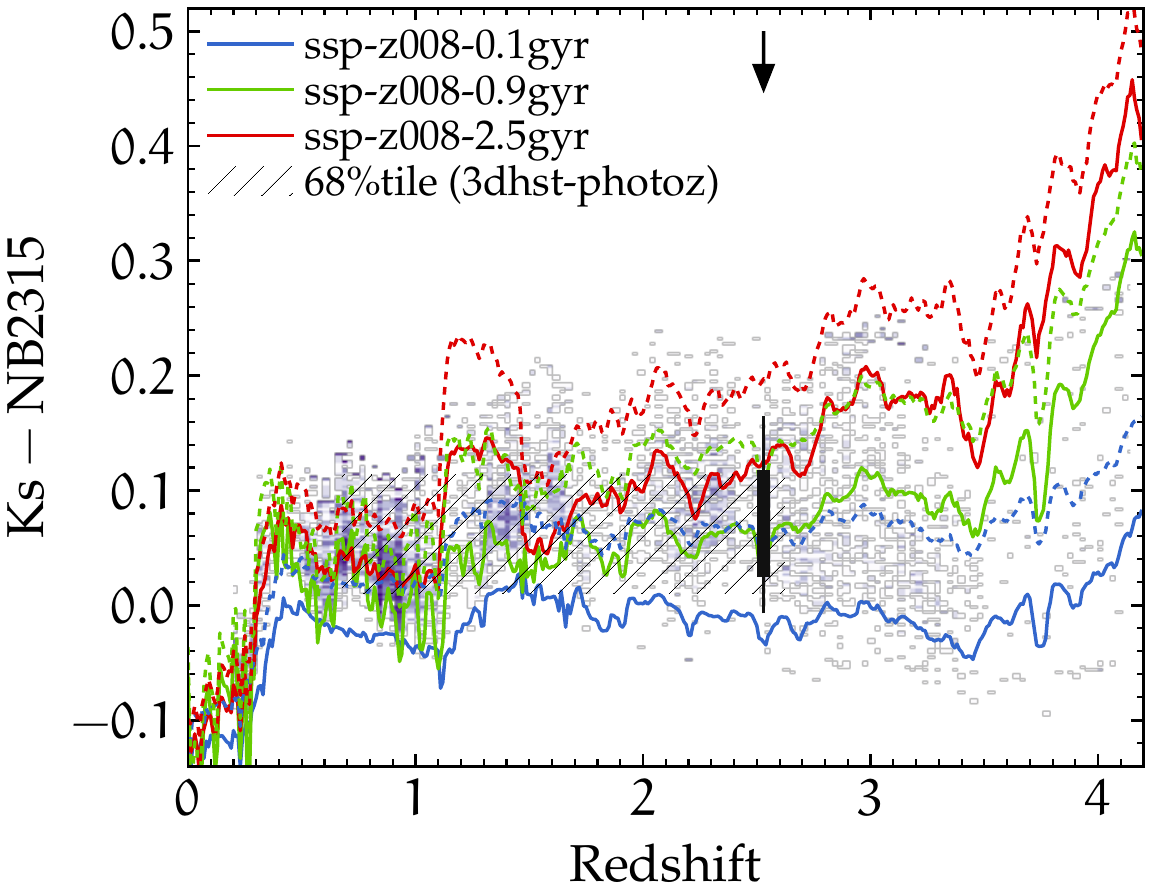}
\caption{Colour term distributions of $K_s-NB2315$ in SSP models 
\citep{Bruzual:2003} with various ages (from the bottom, 0.1 Gyr: {\it blue}, 0.9 
Gyr: {\it green}, and 2.5 Gyr: {\it red}). Solid and dashed curves assume zero 
extinction and $A_V=1$, respectively. The metallicities are fixed to $Z=0.008$. 
The black arrow indicates the narrowband redshift for \ha\ line ($z=2.53$). The 
black hatch region indicates the 68th percentile (\%tile) colour term distribution 
inferred from the SED of the photo-$z$ sources at $K_s=$ 21--24 mag from the 3D-HST 
\citep{Skelton:2014}. Black thick and thin vertical lines indicate 68th \% and 95th 
\%tiles of the colour-term distributions of HAEs inferred from the derived SEDs by 
the {\sc fast} SED fitting (\S3.2).} 
\label{fig14}
\end{figure}

Prior to the simulation, we estimated colour term variations of $K_s-NB2315$ due to 
the difference of filter wavelength to introduce the intrinsic colour term errors 
for the simulation. Derivation of the typical colour term is important also to 
minimise the systematic error of derived narrowband flux and EW for our sample. 
Figure~\ref{fig14} shows colour term distributions across redshift between 0 and 4 
inferred from spectral energy distributions (SED) of single stellar population (SSP) 
synthesis models \citep{Bruzual:2003} with ages of 0.1, 0.9, and 2.5 Gyrs. These 
stellar metallicities are fixed to $Z=0.008$ with two different extinctions 
($A_V=0$ and 1). Also, the figure shows the colour terms of photometric redshift 
sources from 3D-HST \citep{Skelton:2014} whose SED spectra are based on the 
{\sc fast} SED-fitting code \citep{Kriek:2009}. The initial parameters in the 
SED-fitting are the same as for the HAE samples when we estimate their physical 
properties (\S\ref{s3.2}). 
We also checked the colour term distribution with other models with different star 
formation histories or initial parameters for the SED-fitting, and confirmed that 
those colour terms do not greatly deviate from the range covered by those shown in 
the figure. We should note that none of these models include nebular emission 
components. We incorporate the colour term variation of $0.06\pm0.005$ inferred 
from photo-$z$ sources at $K_s=$ 21--24 mag (Fig.~\ref{fig14}) in our simulation 
for the colour completeness, which is broadly consistent with that of derived HAE 
sample (Fig.~\ref{fig14}). 

We then conduct a Monte Carlo simulation similarly to the detection completeness, 
but we embed the PSF into the $K_s$ band image as well as $NB2315$ image. 
We allow for random variation in the $K_s$ magnitudes following the colour term 
distribution noted above. Moreover, we incorporate additional line flux to each 
PSF model for the narrowband image. We embedded a \ha\ line flux from the 
narrowband magnitude using the fitted power-law regression (shown in the lower 
panel of Fig.~\ref{fig13}). Since narrowband magnitudes and derived narrowband 
fluxes of narrowband emitters are tightly correlated, this assumption would be 
reasonable. 
However, we should note that narrowband emitters with lower line fluxes should have 
lower completeness for a given narrowband magnitude. This could cause a systematic 
error for the analyses of distribution functions (\S3.4) at the faint end. Based on 
recovered $NB$ and $K_s$-band photometry derived by the double image mode of the 
SExtractor, we calculate the recovery rate by counting objects which satisfy the 
narrowband selection criteria (eq.~\ref{eq1} and \ref{eq2}). 

The final recovery rate combining the detection completeness with the selection 
completeness is shown in Fig.~\ref{fig13}. We fit the result by an arctangent 
function, 
\begin{equation}
C(NB) = 0.426 - 0.357 \times Arctan(4.047\times(NB - 23.484)), \label{eq11}
\end{equation}
which gives a better solution than polynomial and Gauss error functions. 
$C(NB)$ expresses the completeness value as a function of the narrowband magnitude 
of the HAE samples. We use this correction function to derive the \ha\ luminosity 
and stellar mass functions in \S\ref{s3.4}. The 68 per cent completeness 
corresponds to a line flux of $1.7\times10^{-17}$ erg~s$^{-1}$cm$^{-2}$ or SFR of 
4.2 M$_\odot$yr$^{-1}$ at $z=2.53$. 

We can also estimate the contamination rate due to photometric errors by using this 
simulation. We run the same simulation without adding line flux to the narrowband 
images, and then calculate the contaminants that show the $K_s-NB2315$ colour 
excess due to those photometric errors including the colour term variation. As a 
result, we find that our selection criteria (eq.~\ref{eq1} and \ref{eq2}) 
effectively ignore such contaminations since the derived contamination rates are 
less than 0.2 per cent.


\section{Effect of filter profile}\label{a3}
Since our target is the protocluster at $z=2.53$, associated members are expected 
to gather around the centre of the system along the line of sight rather than 
homogeneously distribute within the redshift slice covered by the narrowband filter. 
Indeed, bright HAEs which have been confirmed by our past spectroscopy at NIR 
\citep{Shimakawa:2014,Shimakawa:2015b} tend to be located around the protocluster 
redshift (see literature or Fig.~\ref{fig16}). 
In this context, the profile of the narrowband filter function should have a 
relatively smaller impact on the measured narrowband fluxes and the survey volume as 
compared to the narrowband imaging for line emitters in the general field. 

However, we should carefully consider the obtained consequences by taking account of 
luminosity dependence on the clustering properties of HAEs as well, which has a 
potential to cause non-ignorable systematic errors in those measurements. 
\citet{Cochrane:2017} have shown that clustering of HAEs should depend on \ha\ 
luminosity in the sense that more luminous HAEs preferentially reside in higher dark 
matter densities. Our dense protocluster sample may be an extreme case of such a 
tight connection between the clustering and \ha\ luminosity, as this work argues 
that bursty HAEs apt to be localised in the dense protocluster's cores. 
Thus, we should discuss the possibility that the intergroup members might be more 
widespread and thus more HAEs are located near the cut-on and cut-off of the filter 
response functions than HAEs in the dense group cores. Since such a systematic error 
may affect our results and conclusion, if any, what follows are a couple of 
simulations that test the effects by the narrowband filter profile on the derivation 
of \ha\ luminosity function and narrowband fluxes of HAEs.


\subsection{Impacts on H$\alpha$ luminosity function}\label{a3.1}

The partially top-hat function of the $NB2315$ narrowband filter essentially probes 
brighter HAEs in wider volume along the line of sight while those at the edge of the 
filter transmission curve can be selected as fainter sources due to the flux loss. As a 
result, observed \ha\ luminosity function can be steeper than the actual one since we 
would tend to unexpectedly count more/less HAEs in the fainter/brighter \ha\ luminosity 
range \citep{Sobral:2009}. This could also produce the lower cutoff \ha\ luminosity in 
the intergroup regions than that of the dense groups as identified by this paper. 

We simulate the impact of this issue by assuming that the intergroup HAEs 
distribute homogeneously along the line of sight. Although this assumption seems to be 
drastic according to the redshift distribution of the existing spec-$z$ sources, 
it should be reasonable to evaluate the effect of the filter profile in such a worst 
possible case. The simulation is made with reference to \citet{Sobral:2009}. We first 
make more than $10^5$ samples with various \ha\ luminosities along the \ha\ luminosity 
function in the group region. We then spread the objects to random redshifts in the 
range of (C) $z=$ 2.51--2.535 or (F) $z=$ 2.49--2.56. The former redshift range 
roughly corresponds to the redshift distribution of HAEs the dense groups by 
\citet{Shimakawa:2014}, and the latter assumes that HAEs are distributed evenly over 
the \ha\ redshift covered by the narrowband filter. After that, we compute the 
recovered \ha\ luminosities depending on those redshifts according to the system 
throughput of the $NB2315$ filter. Based on the recovered \ha\ luminosities of the 
sample, we can reconstruct the \ha\ luminosity functions in the two cases. One should 
note that we here ignore the spatial dependence of filter response functions within 
the MOIRCS field of view \citep{Tanaka:2011}. Furthermore, while we incorporate 
atmospheric transmission into the filter throughput, the filter profile still would 
not be demonstrated perfectly due to its dependence on such as temperature and angle 
of incidence.

\begin{figure}
\centering
\includegraphics[width=.9\columnwidth]{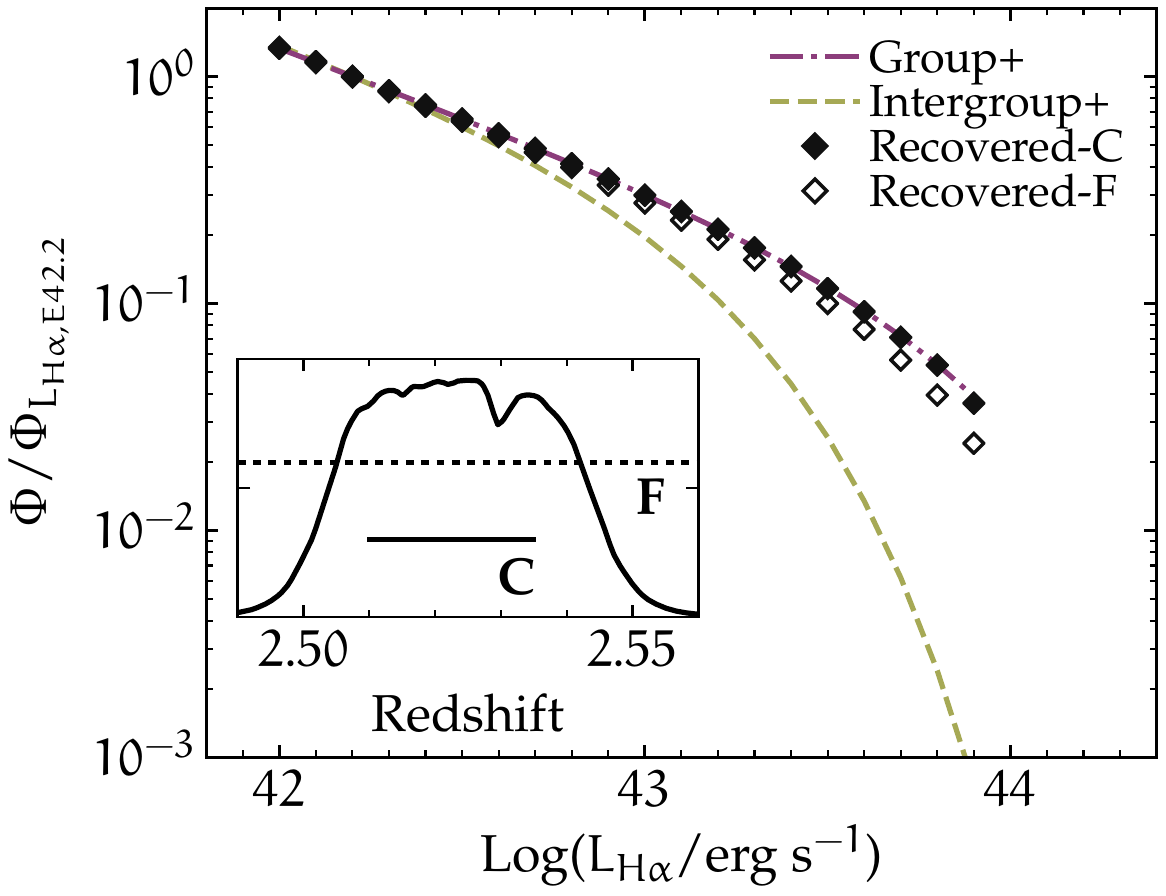}
\caption{\ha\ luminosity functions scaled at $L_\mathrm{H\alpha}=10^{42.2}$ 
erg~s$^{-1}$ (same as the left bottom panel in Fig.~\ref{fig9}). The purple dash-dot 
and yellow dashed curves show the scaled \ha\ luminosity functions in the group and 
the intergroup regions. Black filled and open squares indicate recovered distribution 
functions for the group HAEs by retributing the sample into the different redshift 
space, at $z=$ 2.51--2.535 (pattern C) and 2.49--2.56 (pattern F), respectively (see 
the inset left below).}
\label{fig15}
\end{figure}

Figure~\ref{fig15} shows the obtained retributed \ha\ luminosity functions for the 
group HAEs. Since the case (C) assumes the redshift distribution in the wavelength 
range where the narrowband filter is close to the top-hat function, the recovered 
functions exactly reproduces the original form. Also, the result claims that the 
non-top-hat filter profile has only a minor effect on the \ha\ luminosity function, 
and more homogeneous redshift distributions of HAEs (F) cannot explain the clear 
gap of \ha\ luminosity functions between different densities at the bright end as 
seen in Fig.~\ref{fig9}.


\subsection{Impacts on the star-forming main sequence}\label{a3.2}

\begin{figure}
\centering
\includegraphics[width=.9\columnwidth]{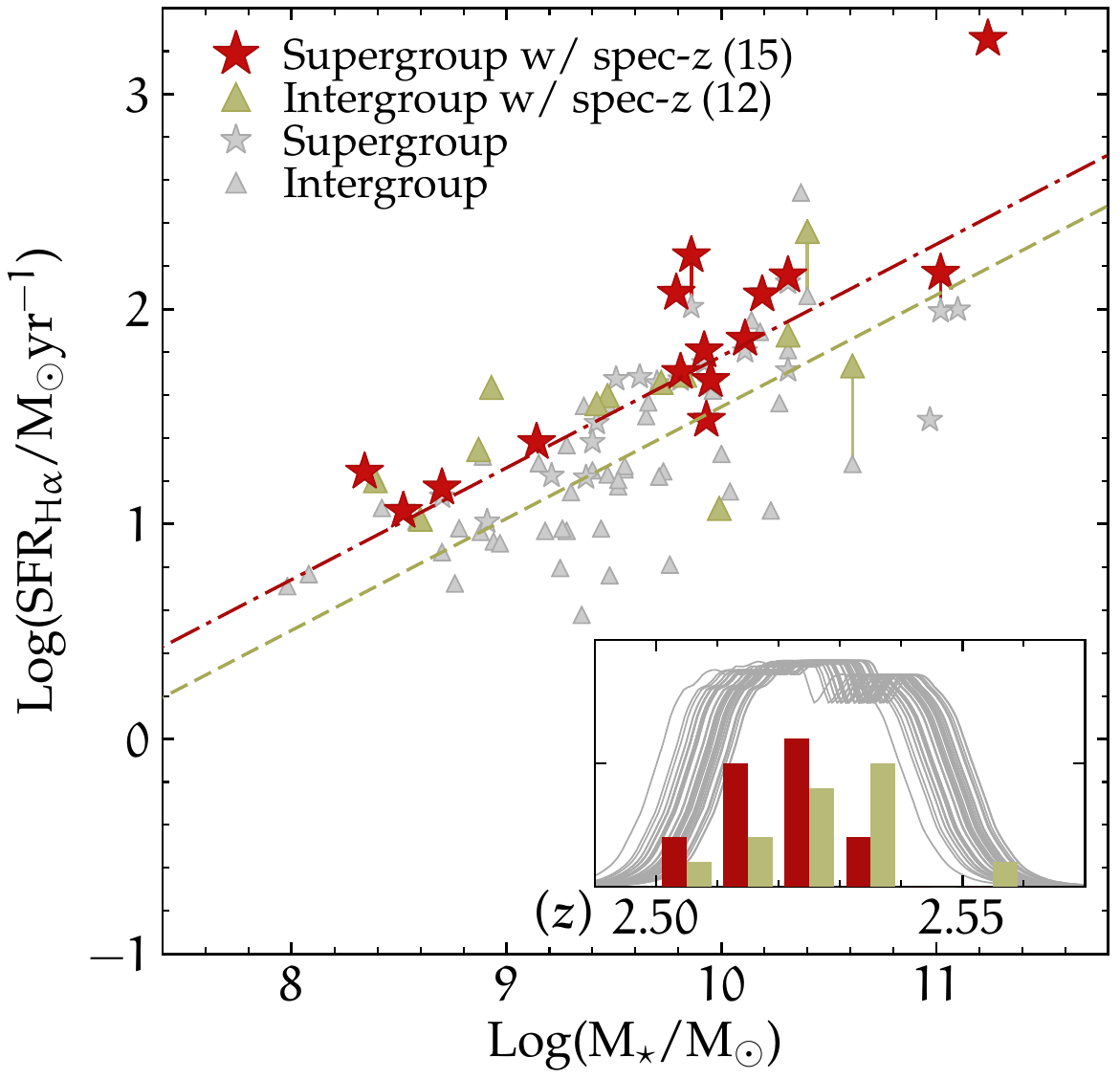}
\caption{The same as Fig.~\ref{fig6}, but we here highlight HAEs with spectroscopic 
confirmations by colours (reds: supergroup HAEs, yellows: intergroup HAEs). We 
correct filter flux losses of spec-$z$ sources where the spatial dependence on the 
filter centre wavelength is corrected. Grey points show the HAEs in each density 
without the correction of the filter flux loss. 
Variation of wavelength shift of the narrowband filter and redshift distributions of 
the HAE samples are shown in the inset right below. The red and yellow histograms 
show HAEs in the supergroup and the intergroup regions, respectively.}
\label{fig16}
\end{figure}

The past spectroscopic follow-up observations \citep{Shimakawa:2014,Shimakawa:2015b} 
do not cover the fainter HAE sources which are newly discovered by our deep 
narrowband imaging reported by this series of papers. Thus, we have few 
spectroscopically confirmed HAEs which appear below the main sequence. 
In fact, Figure~\ref{fig16} indicates that out past NIR spectroscopy mostly follows 
active HAEs, prevent us from investigating the difference of SFR distributions of 
HAEs in various local overdensities. In Fig.~\ref{fig16}, we correct those flux loss 
due to the filter transmittance, which includes the correction of spatial dependence 
of the filter centre wavelength. Except for four HAEs at the filter edges, the 
measurement errors due to the filter profile for HAEs are smaller than 0.1 dex, 
suggesting that the flux loss in the narrowband filter would be a minor effect on 
our results and conclusions.

Besides, we carry out another simulation to check this issue even more carefully. If 
a larger number of HAEs in the intergroup regions were affected by the significant 
\ha\ flux loss relative to those in the dense group regions, it might lead to the 
relatively lower SFR estimate selectively for galaxies in intergroup regions (as 
seen in Fig.~\ref{fig6}). To test such a possibility, we retribute observed 
narrowband flux distribution of the supergroup HAEs and then compare them with those 
in the intergroup HAEs by assuming the different redshift distribution as conducted 
in Appendix \ref{a3.1}. 

\begin{figure}
\centering
\includegraphics[width=.9\columnwidth]{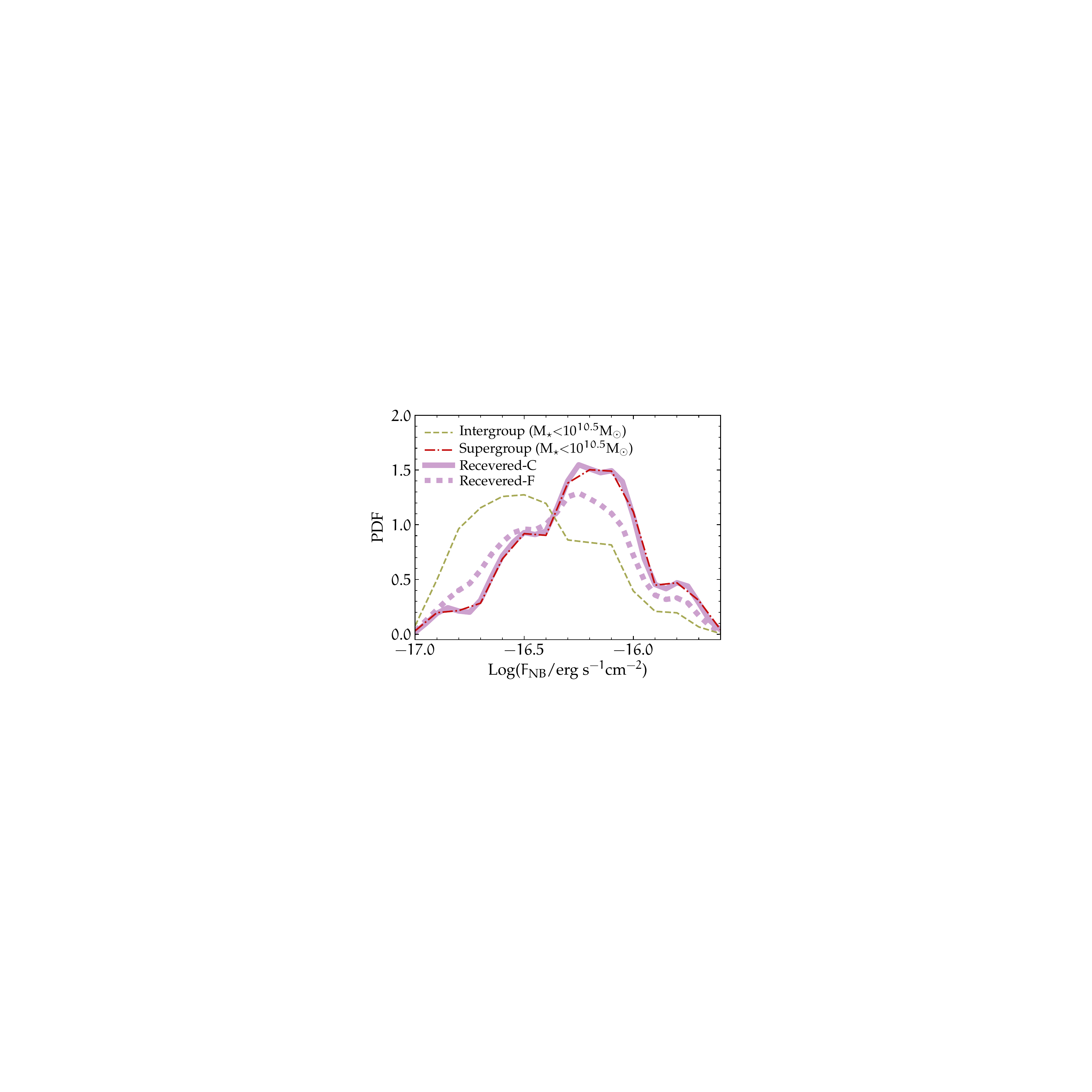}
\caption{Probability density distributions (PDFs) of narrowband fluxes of HAEs in 
the supergroup (red dash-dot) and the intergroup (yellow dashed) regions. Purple 
solid and dotted curves are recovered PDFs by assuming the uniform redshift 
distribution (C) $z=$ 2.51--2.535 and (F) 2.49--2.56, respectively. We employ only 
HAEs at the stellar mass lower than $10^{10.5}$ \msun\ to minimise the effects from 
the higher cutoff in the stellar mass function of the group HAEs.}
\label{fig17}
\end{figure}

Figure~\ref{fig17} demonstrates how more widespread redshift distribution affects 
the narrowband flux distribution. We employ only HAEs with stellar masses lower than 
$10^{10.5}$ \msun\ where we do not see a clear difference in the stellar mass 
distribution between the group HAEs and the intergroup HAEs. The assumption C 
($z=$ 2.51--2.535) well recovers the obtained narrowband distribution of HAEs in the 
supergroup region, indicating that the filter flux loss should be ignorable at this 
redshift range. On the other hand, we see a slight difference in the case of more 
homogeneous redshift distribution over the narrowband filter (F). Since the case F 
presumes that galaxies are homogeneously located at the wider redshift range even 
in which the filter flux loss is significant. The objects with significant flux 
losses are recovered as fainter sources, and then those narrowband flux distribution 
is slightly shifted towards the fainter side. However, the margin of flux 
distribution between supergroup and intergroup HAEs seems to 
remain consistent, suggesting that the only filter profile cannot explain the 
enhanced star-forming activities in the dense cores. However, these results also 
note that we cannot fully ignore this error as some sources show significant flux 
losses due to the non-top-hat function of the narrowband filter. 

Lastly, one may wonder about the possibility that we are missing fainter HAEs in 
higher-densities. We stress that the limiting magnitudes of $NB2315$ and $K_s$ bands 
have only variations of smaller than 0.1 mag over the field of view, and we do not 
see a clear difference of sky background in source photometries of HAEs with the 
SExtractor as a function of the mean projected separation. Indeed, we reconstruct 
the object masks based on the combined science images by the one-stage process, and 
thus such errors from over-subtraction of sky background can be minimised. These 
suggest that spatial dependence of the sampling completeness should be ignorable. 
Thus, at least narrowband flux distribution in high densities should not be caused 
by observational bias. Furthermore, we cannot identify the prominent environmental 
dependence of the \ha\ luminosity function at the faint end. Considering all the 
various factors together, we conclude that enhanced star formation of HAEs in the 
supergroup regions is a more likely intrinsic trend rather than those resulting from 
any systematic error.

\bsp	
\label{lastpage}
\end{document}